%% file: ms.tex
\documentclass[]{aastex631}

\newcommand{\rev}[1]{\textnormal{#1}}

\newcommand{\heasarc}{\textit{HEASARC}}

\newcommand{\chandra}{{\it CHANDRA}}
\newcommand{\rxte}{{\it RXTE}}
\newcommand{\asca}{{\it ASCA}}
\newcommand{\rosat}{{\it ROSAT}}

\newcommand{\exosat}{\textit{EXOSAT}}

\newcommand{\suzaku}{{\it Suzaku}}
\newcommand{\xmm}{{\it XMM-Newton}}

\newcommand{\ec}{$\eta$~Car}
\newcommand{\wro}{WR~140}

\newcommand{\swift}{{\it Swift}}
\newcommand{\nicer}{{\it NICER}}

\newcommand{\ligo}{{\it LIGO}}

\newcommand{\msun}{M$_{\odot}$}

\newcommand{\rsun}{$R_{\odot}$}

\newcommand{\kms}{\mbox{km~s$^{-1}$}}
\newcommand{\ks}{\mbox{km~s$^{-1}$}}
\newcommand{\fluxcgs}{ergs~cm$^{-2}$~s$^{-1}$}
\newcommand{\lumcgs}{ergs~s$^{-1}$}

\newcommand{\CIII}{C~III}
\newcommand{\degree}{\mbox{$^{\circ}$}}

\usepackage{hyperref}

\usepackage{natbib}
\usepackage{epstopdf}
\usepackage{rotating}

\shorttitle{X-ray Monitoring Observations of \wro}
\shortauthors{Pollock et al.}

\begin{document}

\title{Competitive X-ray and Optical Cooling in the Collisionless Shocks of WR~140}

\author{A.~M.~T.~Pollock}
\affiliation{Department of Physics and Astronomy,
University of Sheffield,
Sheffield S3 7RH, UK}

\author{M.~F.~Corcoran}
\affiliation{CRESST and X-ray Astrophysics Laboratory, NASA/Goddard Space Flight Center, Greenbelt, MD 20771, USA}
\affiliation{The Catholic University of America, 620 Michigan Ave., N.E. Washington, DC 20064, USA}

\author{I.~R.~Stevens}
\affiliation{School of Physics and Astronomy, University of Birmingham, Edgbaston, Birmingham, B15 2TT, UK}

\author{C.~M.~P.~Russell}
\affiliation{The Catholic University of America, 620 Michigan Ave., N.E. Washington, DC 20064, USA}

\author{K.~Hamaguchi}
\affiliation{CRESST and X-ray Astrophysics Laboratory, NASA/Goddard Space Flight Center, Greenbelt, MD 20771, USA}
\affiliation{Department of Physics, University of Maryland, Baltimore County, 1000 Hilltop Circle, Baltimore, MD 21250, USA}

\author{P.~M.~Williams}
\affiliation{Institute for Astronomy, University of Edinburgh, Royal Observatory, Edinburgh EH9 3HJ, UK}

\author{A.~F.~J.~Moffat}
\affiliation{D\'epartement de physique and Centre de Recherche en Astrophysique du Qu\'ebec (CRAQ), Universit\'e de Montr\'eal, C.P. 6128, Succ.~Centre-Ville, Montr\'eal, Qu\'ebec, H3C 3J7, Canada}

\author{G. Weigelt}
\affiliation{Max-Planck-Institut f\"ur Radioastronomie, Auf dem H\"ugel 69, 53121 Bonn, Germany}

\author{V.~Shenavrin}
\affiliation{Sternberg Astronomical Institute, Moscow State University, Universitetskii pr. 13, Moscow, 119992 Russia}

\author{N.~D.~Richardson}
\affiliation{Embry-Riddle Aeronautical University, Physics and Astronomy Department,  3700 Willow Creek Road, Prescott, AZ 86301, USA}



 \author{D.~Espinoza} 
\affiliation{The Catholic University of America, 620 Michigan Ave., N.E. Washington, DC 20064, USA}

 \author{S.~A.~Drake}
 \affiliation{CRESST and X-ray Astrophysics Laboratory, NASA/Goddard Space Flight Center, Greenbelt, MD 20771, USA}
 \altaffiliation{Emeritus}

\begin{abstract} 
\wro\ is a long-period, highly eccentric Wolf-Rayet star binary system with exceptionally well-determined orbital and stellar parameters. Bright, variable X-ray emission is generated in shocks produced by the collision of the winds of the WC7pd+O5.5fc component stars.  We discuss the variations in the context of the colliding-wind model using broad-band spectrometry from the \rxte, \swift, and \nicer\ observatories obtained over 20 years and nearly 1000 observations through 3 consecutive 7.94-year orbits including 3 periastron passages. The X-ray luminosity varies as expected with the inverse of the stellar separation over most of the orbit: departures near periastron are produced when cooling shifts to excess optical emission in \ion{C}{3}~$\lambda5696$ in particular. We use X-ray absorption to estimate mass-loss rates for both stars and to constrain the system morphology. The absorption maximum coincides closely with inferior conjunction of the WC star and provides evidence of the ion-reflection mechanism that underlies the formation of collisionless shocks governed by magnetic fields probably generated by the Weibel instability. Comparisons with K-band emission and \ion{He}{1}~$\lambda$10830 absorption show that both are correlated after periastron with the asymmetric X-ray absorption. Dust appears within a few days of periastron suggesting formation within shocked gas near the stagnation point. X-ray flares seen in \ec\ have not occurred in \wro, suggesting the absence of large-scale wind inhomogeneities. Relatively constant soft emission revealed during the X-ray minimum is probably not from recombining plasma entrained in outflowing shocked gas.
\end{abstract}


\section{\wro: A Colliding-Wind Binary Laboratory}

Massive stars, though extremely rare in the local Universe, are important cosmic engines which generate those chemicals critical for planetary %
habitability. There are many open questions about how these unusual stars evolve and die, mostly having to do with mass and angular momentum loss during their short lives
\citep[e.g.][]{2003A&A...404..975M}, %
and how such losses depends on the (evolving) underlying stellar parameters of effective gravity, temperature and chemical abundance, and on initial rotation rate and other complicating factors. %
Additionally, %
the majority of massive stars either have binary companions or were part of a multiple system at some point in their history \citep{Sana:2012fk}, so that binary-driven mass exchange and loss can play an important evolutionary role \citep{Vanbeveren:1979hc,2008MNRAS.384.1109E}, and  may play a key role in forming massive gamma-ray burst progenitors by enhancing the stellar rotation rate \citep{2010NewAR..54..206L}. Massive binaries are the progenitors of the gravitational wave mergers seen by \ligo, but the evolutionary pathway between progenitor massive binary systems and \ligo\ sources is uncertain, and an area of current interest \citep[see, for example,][]{2018MNRAS.481.1908K}.

Observations of massive binaries %
serve as the touchstone for models of stellar structure and evolution in the upper Hertzsprung-Russell Diagram, since the mass and radius of stars in binary systems can be directly measured using simple Newtonian mechanics and Euclidean geometry,  as long as the observer is fortuitously oriented with respect to the orbital plane of the system, or clever enough.  
Massive binaries provide an observational connection between the underlying stellar parameters and stellar wind mass loss \citep[for example,][]{1990ASPC....7..294C}. %
Wind properties (densities and terminal velocities) have %
been determined %
using
radio measures of thermal free-free 
emission \citep{Wright:1975pb}; analysis of UV spectra of P-Cygni profiles in resonance lines \citep{Snow:1976rf};  and through measures of broad optical emission lines \citep{Leitherer:1988yu}. However, each of these methods carries its own assumptions, biases, and uncertainties.

Study of the orbital dependence of the X-ray emission from massive colliding-wind binaries is an independent and increasingly important means for constraining the properties of stellar wind mass-loss. This is because the localized ``bow-shock'' created where the winds from massive binary companions collide %
produces high-temperature gas which emits in the thermal X-ray band (0.1--10 keV). %
Because the X-ray emission occurs near the apex of the colliding-wind shock, it can be used as a probe of local conditions of the stellar winds in situ. The %
X-ray emission measures, temperatures and absorption column densities derived from analysis of the X-ray spectrum as a function of orbital phase provide unique
measures of the dependence with distance from the photospheres of the densities and speeds of the stellar winds. Theoretical modeling of X-rays produced by colliding-wind shocks dates back more than forty years \citep{1976SvA....20....2P, 1976SvAL....2..138C} and has remained an active area of research as computational resources and numerical models have become more sophisticated
\citep[e.g.][]{Stevens:1992yu, usov92, 2007ApJ...660L.141P, 2010MNRAS.403.1657P, Parkin:2011lr, Russell:2013qf}.

The small number of highly eccentric, long-period WR+O binaries are a particularly important means to establish the connection between stellar wind properties and stellar parameters through X-ray variability studies. Such systems offer a number of observational advantages.  Since the stellar separation varies as the stars pass from apastron to periastron and back again, the emission measure of the shocked material along the wind collision boundary varies with orbital phase and constrains the spatial wind density distribution from apastron through periastron passage and back again. X-ray temperatures %
constrain the local stellar wind velocities of the upstream pre-shocked gas near the point of impact as a function of stellar separation. Variations in X-ray absorption, produced by the unshocked stellar winds as the stars orbit each other, constrain mass loss rates and the geometry of both the shocked and unshocked winds.  In addition, because the stars are well separated for most of the orbit, mass transfer might only  occur now near periastron, although a modest amount has probably taken place earlier over the course of \wro's evolution \citep{2021arXiv210110563T}.

The outstanding example of an eccentric, long-period, X-ray-variable colliding-wind binary is undoubtedly \wro\ (HD~193793), a WC7pd+O5.5fc binary of period $7.94$ years and eccentricity $0.900$. 
Its exceptional nature was first recognized via periodic dust formation events \citep{1978MNRAS.185..467W, 1987IAUS..122..453W, 1990MNRAS.243..662W, Williams:1994vn, 1996RMxAC...5...47W, Williams:2009rf}.  This periodic outpouring of dust served as the ``smoking gun'' to establish the system's long-period and extreme eccentricity which is one of the highest of all known stellar binary systems\footnote{There are 38 binary systems with orbital eccentricity$>0.89$ in the 9th Catalog of Spectroscopic Binaries \citep{2004A&A...424..727P} of which \wro\ is the only WR binary.  Worth noting, the peculiar massive long-period binary \ec\ also has a high orbital eccentricity \citep[near 0.9, ][though it has not been directly measured]{Madura:2012qf}.  The catalogued system with the highest eccentricity is the low-mass binary 
HD~137763 ($P=889.95$~days, $e=0.976$).}). 
Its orbit and the properties of both stars have been well established from ground-based observations
\citep{2005ApJ...623..447D, Monnier:2011qy, Fahed:2011qy}.
New analysis by \cite{2021arXiv210110563T}, hereafter T21, using optical interferometry from the Center for High Angular Resolution Astronomy (CHARA) Array, combined with archival interferometry and new and archival stellar spectra, has provided precise values for the orbital elements and stellar masses. 
As a result of these detailed studies,
\wro\ has
the best established stellar and orbital parameters of any WR-type binary, rivalling other very high-eccentricity systems of assorted spectral type discussed, for example, by \citet{2018AJ....156..144F}. 

In a wider context, the shocks generated by the wind interactions in massive stellar binary systems and \wro\ in particular \citep{2005ApJ...629..482P} are prime and arguably optimal exemplars of the physics of collisionless shocks.
Collisionless shocks play important roles throughout the Universe in the dynamics of plasmas across the widest
range of settings \citep[e.g.][]{2016RPPh...79d6901M,2019SciA....5.9926D}
from the flow of the solar wind into the Earth's magnetosphere and later into the heliopause;
through supernova remnants blasting into the interstellar medium;
to merging clusters of galaxies.
The behaviour of all these systems is governed by interlocking varieties of physics that encompass one or more of
shock formation, development and dissipation;
independent ion and electron heating and cooling, excitation and emission;
non-thermal particle acceleration;
dust formation; 
magnetic field generation and reconnection;
plasma waves;
turbulence;
and dynamical instabilities.
\wro\ exhibits many of these in tightly constrained conditions.
It is bright, nearby and unobscured enough to offer observers unrivalled opportunities for the remote study of
shock physics to contrast and complement %
measurements made in solar system space and laboratories on earth. 
By virtue of the combination of this precision with its exceptional brightness, X-ray emission from \wro\ currently serves as a uniquely valuable resource for the study of the behavior of collisionless astrophysical shocks under predictable, repeatable, time-dependent physical conditions.

\wro\ was first detected as a luminous X-ray source on 1984 May 19 by the \exosat\ observatory \citep{1985SSRv...40...63P}, at what subsequently proved to be orbital phase 
$0.889$, and has been observed many times since by X-ray observatories like \rosat\ \citep{1995IAUS..163..512P}, \asca\ \citep{1994PASJ...46L..93K, 2000ApJ...538..808Z}, \xmm\ \citep{De-Becker:2011lr}, \chandra\ \citep{2005ApJ...629..482P}
and \suzaku\ \citep{Sugawara:2015eu}.
\wro\ is the brightest X-ray source among all single or binary massive
stars, rivaled only by \ec\ with which it shares a similar range of observed count rates in many instruments. While the X-ray brightness of \wro\ has obvious advantages, at the same time
care is required regarding the choice of observing modes in order to avoid where possible adverse instrumental effects such as photon pile-up.

Here we report detailed X-ray monitoring of the system from observations with \rxte, \swift, and \nicer, emphasizing the broad-band variations revealed as the stars revolve in their orbits, on timescales of days to weeks, over 3~orbital cycles according to the orbital and stellar parameters given in Table~\ref{tab:params}. Most have been accurately determined through observations except perhaps the more uncertain O-star
\rev{mass loss, for which we have adopted the preferred smallest}
of the clumping-free rates reported by \citet{2006MNRAS.372..801P}. Also given, to inform the discussion below, are shock-cone opening angles, $\theta_S$, and the shock stand-off distance from the WC7 star
\rev{as a fraction of the binary separation, $D$. These were derived from analytical geometrical models of radiative and adiabatic binary shock formation \citep{1996ApJ...469..729C, 2009ApJ...703...89G}
and depend on the momentum flux ratio, $\eta$. Their rigor and applicability were assessed by \citet{2018MNRAS.477.5640P} based on 2D numerical fluid simulations.}

We investigate variations of the emergent X-rays in the intensity and spectrum of the shock emission as modulated by photoelectric absorption during transfer through cooler pre-shock wind and the interstellar medium, on timescales
according to orbital phase from
hours and days to years and beyond, over the three most recent orbital cycles that span the first 20 years of the 21st century.
As expected, in common with longer-wavelength regimes, X-ray events move much more quickly near periastron than at other phases, so we requested observations at increasingly higher cadence in order
to sample appropriate details of the system's emerging behaviour when the stars are close. Ultimately { nearly a thousand} separate observations have been accumulated {and analyzed}.

In \wro's eccentric 7.94-year orbit, %
the succession of Keplerian events of geometrical relevance
occur within a few months of periastron:
the O star is in front at $-129$~days;
followed by quadrature when the stars are sideways on with the WC star moving towards the observer at $-10$ days;
the WC star in front at $+9$~days;
and quadrature with the WC star moving away at $+104$~days.
These times are accurate to about a day near periastron and a week otherwise.
Between periastron and apastron, the distance between the companion stars varies from 1.5 to 27.5 AU, 
similar to the separations of %
Mars and Neptune from the Sun.

The paper is organized as follows: Section \ref{sec:obs} presents an overview of the X-ray satellite observatories, \rxte, \swift, and \nicer, and their work on \wro.  In Section~\ref{sec:specmodel} we derive a suitable spectral model for \wro\ which we then use consistently to model the spectra from all three observatories.  In Section~\ref{sec:xlc}, we present the full X-ray lightcurve from 2001 through 2020, and %
 {discuss} %
the variations of the 
spectral parameters as a function of time and orbital phase
paying special attention in Section~\ref{sec:ciii} to the mechanisms of shock formation and cooling in \wro.
Section \ref{sec:discuss} discusses %
the 
implications for the understanding of the mass loss from the WR and O components, and we summarize the main results in Section \ref{sec:conc}.

\begin{deluxetable}{lrl}
\tabletypesize{\scriptsize}
\tablecaption{Adopted Orbital and Stellar Parameters\label{tab:params}}
\tablehead{\colhead{Parameter} & \colhead{Value} & \colhead{Reference or Comment}}
\startdata
Period (days)  &     2895$\pm0.17$  & T21   (from Table 3,  ``Adopted Fit'') \\
eccentricity  &    0.9883 $\pm$  0.0013 & T21   \\
Time of Periastron Passage (MJD) &   60637.23 $\pm$ 0.53   & Best fit $T_{o}$ from T21     \\
$\omega_{\mbox{WR}}$ (degrees) &     227.44  $\pm$ 0.52  & T21     \\
$\Omega$ (degrees)  &  353.87   $\pm$ 0.67 & T21   \\
$i$ (degrees)  & 119.07  & T21    \\
$a$ (AU) & 13.55$\pm$ 0.21 & T21 \\
$M_{WR}$ (\msun)   &    10.31 $\pm$   0.45 & T21   \\
$M_{O}$ (\msun)  &     29.27 $\pm$    1.14 & T21   \\
$\dot{M}_{WR}$ ($10^{-6}$\msun~yr$^{-1}$) & 43.0  & \cite{2006MNRAS.372..801P} \\
$\dot{M}_{O}$ ($10^{-6}$\msun~yr$^{-1}$) & 0.8  &  \cite{2006MNRAS.372..801P} \\
$V_{\infty, WR}$ (km~s$^{-1}$) &2860  &\cite{2001aap...376..460S} \\
$V_{\infty, O}$ (km~s$^{-1}$) &  3100  & \cite{2001aap...376..460S} \\
$\eta$ & 49.6 & $=(\dot{M}_{WR}V_{\infty, WR}$)/($\dot{M}_{O}V_{\infty, O}$)\\
Adiabatic-Shock Opening Angle (degrees)\tablenotemark{a} & 41.3 &  \cite{2009ApJ...703...89G} equation (11) \\
Radiative-Shock Opening Angle (degrees)\tablenotemark{a} & 31.7 & \cite{1996ApJ...469..729C} equation (28) \\
WR-Shock Stand-off Distance\tablenotemark{b} & 0.876 & \cite{1996ApJ...469..729C} equation (27) \\
Distance (pc)  &   1518 & Gaia~DR2 \& T21 \\
$v$ & $7.07$ & \cite{1984ApJ...281..789M} \\
$b-v$ & $+0.27$ & \cite{1984ApJ...281..789M} \\
$N_{H}$ ($10^{21}$~cm$^{-2})$ & 4.3 & Interstellar absorption column density
\enddata
\tablenotetext{a}{Theoretical asymptotic opening half-angle determined by $\eta$}
\tablenotetext{b}{Fraction of binary separation determined by $\eta$}
\end{deluxetable}

\section{The Observations}
\label{sec:obs}

There are a total of 940
observations of \wro\ by \rxte, \swift\ or \nicer, for a total exposure of 1342.0
ksec, included here. Table~\ref{tab:sumtab} summarizes the observations for each of the missions.

\subsection{\rxte}

The \textit{Rossi X-ray Timing Explorer} \citep[\rxte,][]{rxte}, which operated between
1995 December and 2012 January, was a satellite observatory with the flexibility to observe variable X-ray emitting cosmic objects on timescales from microseconds to years. The primary instrument, the Proportional Counter Array \citep[PCA, ][]{Jahoda:2006lr,2012ApJ...757..159S}, was an array of collimated, non-imaging proportional counters with a field-of-view of about $1^{\circ}$, sensitive to X-rays emitted in the $\sim3-60$~keV band.  This band includes a significant fraction of %
the absorbed emission generated within \wro's wind-wind collision shock which falls almost entirely between $0.5$ and $10$~keV. However, the lack of soft response below 3~keV made \rxte\  insensitive to column density variations at relatively low values of $N_{H}\lesssim10^{22}$~cm$^{-2}$. Also, \rxte's wide field-of-view means that PCA spectra are contaminated by cosmic background at a level dependent on satellite roll angle. \rxte\ had limited spectral resolution  ($E/\Delta E$ = 3.45 - 6.15) between 3 and 8~keV, too coarse for studies of emission lines. 

The first observation of \wro\ by \rxte\ was obtained on
2000 December 09,
63 days before \wro's periastron passage (which occurred on 2001 February 10), after
which \rxte\ observed the system approximately once per week
for just over  {two} years.
After a  {two} year hiatus starting in 2003 March,
observations resumed on 2005 March 08,
39 days after apastron ($\phi=$ 0.513), 
and continued %
for  {six} years
until 2011 December 23, just before end of mission, at a variable cadence of a few observations per month to one per day during the periastron passage
and X-ray minimum in 2009 January.
As for many space observatories, \rxte\ observing time
was awarded on an annual basis so that its cumulative
\wro\ campaign comprised 8 separate proposals after being initiated to inform the context of  {two} \chandra\ observations of \wro\ being made at the time
and whose success they helped guarantee. These data have been discussed previously \citep{Corcoran:2011qv, Pollock:2012qv}.  Most recently, \cite{2021MNRAS.500.4837Z} analyzed these data (along with \chandra\ X-ray spectra) and suggested that the stellar winds are significantly clumped and/or non-spherically symmetric.

Though the PCA consists of 5 individual Proportional Counter Units (PCUs, conventionally numbered 0--4), we concentrate here on data from PCU-2, which provides the most complete and reliable data, and in particular, the data obtained in Layer 1 of PCU-2, where most of the X-rays from a relatively soft source like \wro\  are measured. We obtained and analyzed these data from the \rxte\ data archive at the HEASARC, NASA's archive of high energy space mission data. For each observation, we created individual responses for and instrumental background estimates of the PCU-2 layer 1 data using software tools (\texttt{pcarsp} and \texttt{pcabackest}) provided by the \rxte\ Guest Observer Facility and distributed with the HEASARC's \texttt{HEASoft} software package.  Note that the background estimates provided by \texttt{pcabackest} provide estimates of the internal detector background and external charged particle background and do not account for cosmic background.  Thus the amount of residual cosmic ``sky'' background included in the $\approx 1^{\circ}$ field of view of %
PCU-2 will be different than the amount of residual cosmic background in imaging observations (such as the \swift\ XRT in photon-counting mode) or non-imaging instruments with narrower fields of view (like \nicer).  Difference in cosmic background in the different instruments  complicates comparison between the various observatories.  %
We used calibrations from the most recent version of the PCA calibration data (version 20120110) obtained from the \heasarc\ \rxte\ Calibration Database (CALDB).

\input{sumtab}

\subsection{\swift}

We used the X-ray Telescope  \citep[XRT,][]{Burrows:2000lq} on the \textit{Neil Gehrels} \swift\ observatory \citep[\swift,][]{2004ApJ...611.1005G} to monitor changes in the X-ray spectrum. %
\swift\ was used to augment the \rxte\ data %
since the XRT's softer instrumental sensitivity and better spectral resolution allow for more precise determination of column densities than possible with \rxte. On the other hand, 
the XRT is prone to light leaks at low energies, which can complicate analysis of the soft part of the X-ray spectrum where the effect of wind absorption is most pronounced.
This contamination, known as optical loading, has been particularly relevant in the treatment of
the soft ``relic'' component \citep[][Section \ref{sec:relic} below]{Sugawara:2015eu} %
revealed during \wro's X-ray minimum near periastron.

The XRT is also susceptible to photon pile-up in imaging (``Photon-Counting'', hereafter PC) mode observations if more than one X-ray photon strikes a pixel of the XRT detector before readout, reducing the observed count rate and distorting the spectrum. The count-rate threshold above which this becomes significant is about $0.5~\mathrm{cts}~\mathrm{s}^{-1}$ and applies to most of \wro's orbit except the
minimum near periastron. This was not understood during the early XRT observations. Though
data analysis methods available to mitigate the problem have been used, the XRT's alternative ``Windowed-Timing'' mode (hereafter WT), was later usually judged preferable. Its faster readout incurs the loss of spatial information in one direction leading to higher backgrounds and more difficult treatment of bad detector columns.

The \swift\ XRT observations of \wro\ started on
\rev{2008 August 11,}
 with sporadic phase coverage.  The last observation we analyze here was taken on
 \rev{2018 April 05}. 
 \swift,  {along with \suzaku\ \citep{Sugawara:2015eu}}, provided critical measurements of the variation in the \wro's soft X-ray spectrum 
near the 2009 periastron passage since
only
\rev{\swift\ and \suzaku\ }
were available to measure the \wro\ X-ray spectrum below 3~keV at that time\footnote{\wro\ violated viewing constraints  {of \xmm}, and mitigation of photon pileup in the \chandra\ ACIS camera would have resulted in prohibitively long exposures. In addition, at that time it was unclear if the \rxte\ mission would be extended beyond February 2009; if not, it would have been up to \swift\ alone to monitor  {the details of} the recovery of \wro's X-ray flux from the X-ray minimum state.}. %

The \swift\ data were downloaded from the \heasarc\ data archive, then reprocessed using the standard pipeline software (\texttt{xrtpipeline}, part of the \swift\ XRTDAS software distributed with the HEASoft Analysis package). 
We calculated individual effective areas for the PC and WT mode data. We visually inspected each observation and defined valid ``good-time'' intervals from extracted lightcurves, avoiding times of large excursions in count rates produced by background fluctuations or other, non-source-related variations.  We then extracted spectra and lightcurves from the X-ray events which fell within these good time intervals.  
For the PC-mode data, source spectra were extracted from a circular
region of radius $1.1'$ centered on \wro.  %
Background was extracted from an annulus between $2.7'$ and $8.3'$, whose inner boundary extends beyond the \swift\ PC-mode 90\% encircled energy radius, and which contained no visible X-ray sources. For each PC-mode spectrum we calculated observation-specific effective areas using the \texttt{xrtmkarf} task distributed with the \swift\ XRTDAS, and used the standard XRT PC-mode response function from version 20170501 of the \swift\ XRT CALDB in the spectral analysis. 

For the WT mode data, neither pileup nor optical loading is a significant issue.  For data in windowed-timing mode, source photons were extracted from a region extending to $1.1'$  around the source in the 1-D spatially compressed image, while background was extracted from an apparently source-free region extending to $1.1'$ in the 1-D spatially compressed image.  Because of the compression in the spatial dimension, the background spectrum in the windowed-timing mode data depends on the 
spacecraft roll angle since cosmic sources and bad detector columns included in the spatially-compressed %
vary with roll angle. Thus the WT mode backgrounds are in general different from the backgrounds %
in photon-counting mode but, from examination of a few selected datasets at different roll angles, we estimate that the variation between background levels in different WT observations is probably less than $5\%$. 
We computed individual, observation-specific effective areas for all the WT-mode data and analyzed the data using the standard WT-mode response function. 
Comparison of net fluxes derived from PC-mode data with those derived from the analysis of net WT-mode data indicates that the net WT-mode data are in general 40\% brighter than the net PC-mode data, presumably due to an under-estimation of background contamination in the WT-mode data.

\subsection{\nicer}

The Neutron Star Interior Composition Explorer \citep[\nicer,][]{2014SPIE.9144E..20A} is an X-ray astronomy facility which was installed on the International Space Station ( {ISS}) in June 2017.  \nicer's primary mission is to obtain high-time resolved X-ray spectrometry in the 0.2--12 keV band with moderate spectral resolution of cosmic X-ray sources, primarily of X-ray binary pulsars. %
\nicer's X-ray Timing Instrument \citep[XTI,][]{Prigozhin:2016lr} is comprised of a co-aligned array of 52 %
Focal Plane Modules, each consisting of a matched pair of X-ray ``concentrator'' (XRC) optics a silicon drift detector (SDD) for readout. Each XRC optic collects X-rays over a roughly 30 arcmin$^{2}$ region of the sky centered on the target of interest in the  0.2--12~keV energy band and concentrates them onto an SDD.  
\nicer's %
combination of large effective area, restricted field of view and broad bandpass in the thermal X-ray range makes it especially
well-suited to observe X-ray variables like long-period colliding wind binaries (and other sources) in addition to X-ray pulsars.  %

The \nicer\ data were extracted from the clean, merged photon events files, using data obtained outside of the South Atlantic Anomaly at sun angles of $>40^{\circ}$ to avoid stray-light contamination, using the NICERDAS software package distributed with HEASoft.  We used standard \nicer\ event cleaning criteria\footnote{see \href{https://heasarc.gsfc.nasa.gov/lheasoft/ftools/headas/nimaketime.html}{the help file for \texttt{nimaketime}}  for a description of the standard data screening criteria.} to clean each observation.  \nicer\ is subject to a varying charged particle environment at the high-inclination ISS orbit, which is most noticeable at high energies, so we extracted lightcurves in the $0.4-10$~keV band for all the \nicer\ observations and visually inspected them to exclude intervals of large rapid increases in count rate produced by the variable charged particle environment. In addition, low-energy events produce a ``noise spike'' at energies $\lesssim0.4$~keV.  Therefore, in our analysis, we extracted and fit spectra in the 0.4--10~keV band, which has good overlap with \swift\ and \rxte\ and excludes the \nicer\ spectral bands most affected by background issues.  For each observation we estimated background  using the  \texttt{nibackgen3C50} tool\footnote{See \url{https://heasarc.gsfc.nasa.gov/docs/nicer/tools/nicer_bkg_est_tools.html}.} { \citep{2021arXiv210509901R} }provided by the \nicer\ Guest Observer Facility, which estimates background from a library of spectra extracted from blank-field observations, using rates at energies beyond the XRC cutoff and other proxies. We subtracted this background estimate from the observed spectrum for each observation before fitting.

\section{Determining a Spectral Template Model}
\label{sec:specmodel}

In order to determine a uniform set of phase-dependent broadband X-ray spectral parameters from the moderate resolution data delivered by \rxte, \swift, and \nicer\ 
we created a template model of the spectrum using high-resolution \chandra\ grating spectrometer data, as these contain the most detailed
information on the combination of lines and continuum. In particular, we chose three \chandra\ ObsIDs, 6286, 5419 and 6287,  obtained over 
a  {two} week period in 2006 March and April, only 14 months (0.15
in orbital phase) after the stars passed apastron (on 2005-01-28).  These spectra show the plasma in its most relaxed condition, and stellar wind absorption at or near its lowest. The total exposure time of the combined spectrum is 141~ks.
Because of the resemblance of the spectrum to thermal emission,
we used a simple combination of plasma models in collisional equilibrium with the same adjustments for velocity broadening and non-solar abundances
and subject to absorption by the interstellar medium and the stellar winds, and used  XSPEC \citep{Dorman:2001bh} to fit the combined set of 12 MEG and HEG first-order spectra.

The model we adopted was selected for its ability to reproduce the morphology of the spectrum, so it does not necessarily represent a realistic description of the actual physics of the plasma
which is more likely to be out of equilibrium as discussed, for example, by \citet{2005ApJ...629..482P}.
In XSPEC notation, the model is \texttt{(TBabs(bvapec$_{1}$ + bvapec$_{2}$ + bvapec$_{3}$)},
where \texttt{TBabs} is the ``Tuebingen-Boulder ISM absorption model'' \citep[][which was used to estimate the combination of
ISM and stellar wind absorption columns to the X-ray source]{2000ApJ...542..914W}, and 
the three \texttt{bvapec} components are non-solar-abundance (\texttt{"v"}), doppler-broadened (\texttt{"b"}) plasma emission models \citep{2001ApJ...556L..91S}\footnote{see \url{http://atomdb.org}.}.
When absorption becomes high as the Wolf-Rayet star moves in front
of the colliding-wind source a weak additional soft component not captured by the template model becomes visible, amounting to less than 0.25\% of the maximum count rate.
Known by \citet{Sugawara:2015eu} as the ``relic'' component, its behavior is discussed below in section~\ref{sec:relic}.

Abundances were held in common for the three thermal emission components and are
\rev{mostly representative of the shocked wind of the chemically-evolved WC7 star with a minor contribution,
amounting to less than 10\% according to \citet{2002A&A...388L..20P}, from the shocked wind of the less chemically-evolved O star.
In order to accommodate this, with hydrogen absent in the WC7 star and replaced by helium,}
the abundances of the \chandra\  emission model were fixed: helium to a high value; carbon to a number fraction $\mathrm{C}/\mathrm{He}=0.4$; and nitrogen to zero.

\begin{figure}[htbp] %
   \centering
   \includegraphics[width=6.5in]{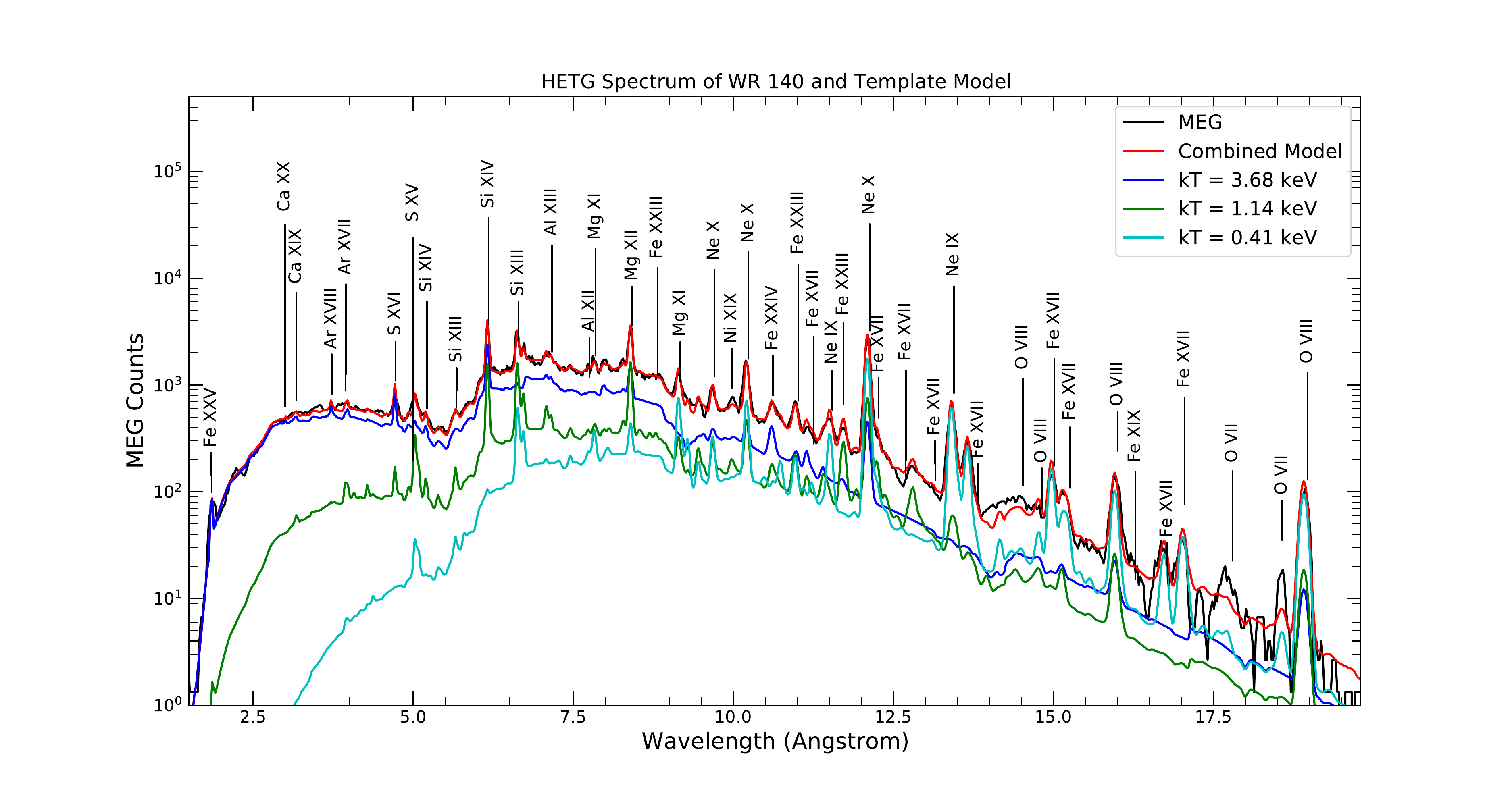} 
   \caption{\chandra\ HETG combined $\pm1$st order MEG spectra from the ObsIDs 6286, 5419, 6287  of \wro\ (black) closest to apastron and the standard template model (red). Strong lines in the 1--20~\AA\ region are marked among which those of \ion{Ne}{10} and \ion{O}{8} are particularly strong.
   }
   \label{fig:tempmo}
\end{figure}

The reference template model parameters from the \chandra\ fit
are given in Table \ref{tab:templ_mo} \rev{including} the 
\rev{best-fit}
abundances used for all \rev{three} emission components, given with respect to solar abundances \citep{1989GeCoA..53..197A} \rev{and as number and mass fractions}.
The best-fit emission measures
of the three thermal components to the HETGS spectra were fixed at values adjusted downwards by an indicative factor of 1.061
calculated from the relative binary separation of the orbit between apastron and the later mean epoch of observation.
With density estimates based on parameters in Table~\ref{tab:params}, these emission measures equate to volumes of less than 1\% of a characteristic volume given by $(1-\cos{\theta_S})D^3$.
Other than the relic component, the luminosity scaling factor was one of the two free parameters of the template model along with the absorbing column density
whose initial value was also thought to be characteristic of low values at apastron or maximum binary separation, although expectations at the outset
were ill-defined of its overall phase dependence. The line-velocity structure was also frozen in the template model, since line velocity broadening is only of relevance at high (grating) resolution.  The HETG spectrum and template model, including the 3 individual emission components representing emission from the shocked gas in the wind collision region, is shown in Figure \ref{fig:tempmo}.

\input{templ_mo.tex}

We then fit this ``standard'' X-ray spectrum model to each individual spectrum from the \rxte, and \swift\ observations, fitting for the overall column densities, using an overall constant to adjust the flux normalization of the model, with the temperatures and relative emission measures of the emission model fixed.  We found that this was adequate to generate acceptable fits to the low-resolution \rxte\ spectra and the relatively low precision \swift\ and \nicer\ spectra (null hypothesis probablilty~$>0.05$) which therefore gave no evidence of any change with orbital phase of the adopted equivalent plasma temperatures although this conclusion is ripe for exploration with high resolution data.

We used an automated Python script to fit the data, using \texttt{pyxspec}, the Python version of the XSPEC spectral analysis package. %
We recorded values of %
all variable parameters, namely those not fixed in Table \ref{tab:templ_mo},
for each spectrum and calculated observed and absorption-corrected fluxes in the 2--10~keV band for all observations. We adopted the 2--10~keV band as standard, since it is consistent with the harder response of \rxte\ that does not extend much below 2~keV and includes the bulk of the colliding-wind emission. Any intrinsic emission from the individual stellar winds is also likely to be mostly confined to the soft band below 2~keV although even this is weak in \wro: the residual soft emission near periastron discussed in Section~\ref{sec:relic} below amounts to less than 2\% of the colliding-wind emission in the same band observed at  {apastron}.

\section{X-ray variability in \wro}
\label{sec:xlc}

\subsection{Cross-Calibration and Cosmic Background}

One important objective of instrumental calibration procedures is to enable stable, robust and consistent estimates of physical parameters. Comparisons of results for calibration standards such as the IACHEC group's work on the supernova remnant 1E~0102.2-7219
\citep{Plucinsky:2017lr}, show that different instruments usually agree to within a few percent with some energy dependence. Similar considerations apply to measurements of \wro.

We found that %
 there were flux offsets between observations from different instruments for observations which overlap in phase and in time.  In addition to the cross-calibration 
uncertainties, flux offsets arise because of the different amount of cosmic X-ray background (diffuse X-ray background along with other possible cosmic sources) included in the  non-imaging observations by \rxte, \nicer, and \swift\ in WT mode, since all these instruments have different fields-of-view in addition to different sensitivities.  The imaging  \swift\ PC-mode data allow the most accurate account for cosmic background since background is directly determined from source-free regions of the PC image.  We therefore corrected the \rxte, \nicer\ and \swift\ WT mode fluxes to the PC fluxes.  We found that we needed to subtract an additional background flux of 0.50
$\times 10^{-11}$~\fluxcgs\ from the \rxte\ 2--10 keV band fluxes, and  0.70
$\times 10^{-11}$~\fluxcgs\ from the \nicer\ fluxes, and 0.80
$\times 10^{-11}$~\fluxcgs\ from the \swift\ WT mode data to bring them into agreement with the \swift\ PC-mode data.  In addition, we needed to divide the background-corrected \rxte\ fluxes by a scale factor of 1.35 to match the \swift\ PC-mode data in order to bring the amplitude of variation of the \rxte\ data into agreement with the \swift\ PC-mode data in the regions near X-ray maximum.
 
\subsection{X-ray Flux Variations in the 2--10 keV Band}
In what follows, times and phases are calculated %
using the following ephemeris:
$$T_{o}(JD)  = 2454846.7270
+ 2895.00
E$$
where  $E$ is the cycle count,  the period is given by T21, and the epoch $T_{o}$
(2454846.7270 = 2009-01-15 05:26:52 UTC)
corresponds to the time of periastron passage from T21, but 2
orbital cycles earlier.  We adopt this value of $T_{o}$ since this periastron passage was  densely monitored by both \rxte\ and \swift.

 \begin{sidewaysfigure}[htb!] %
   \centering
    \includegraphics[width=10in]{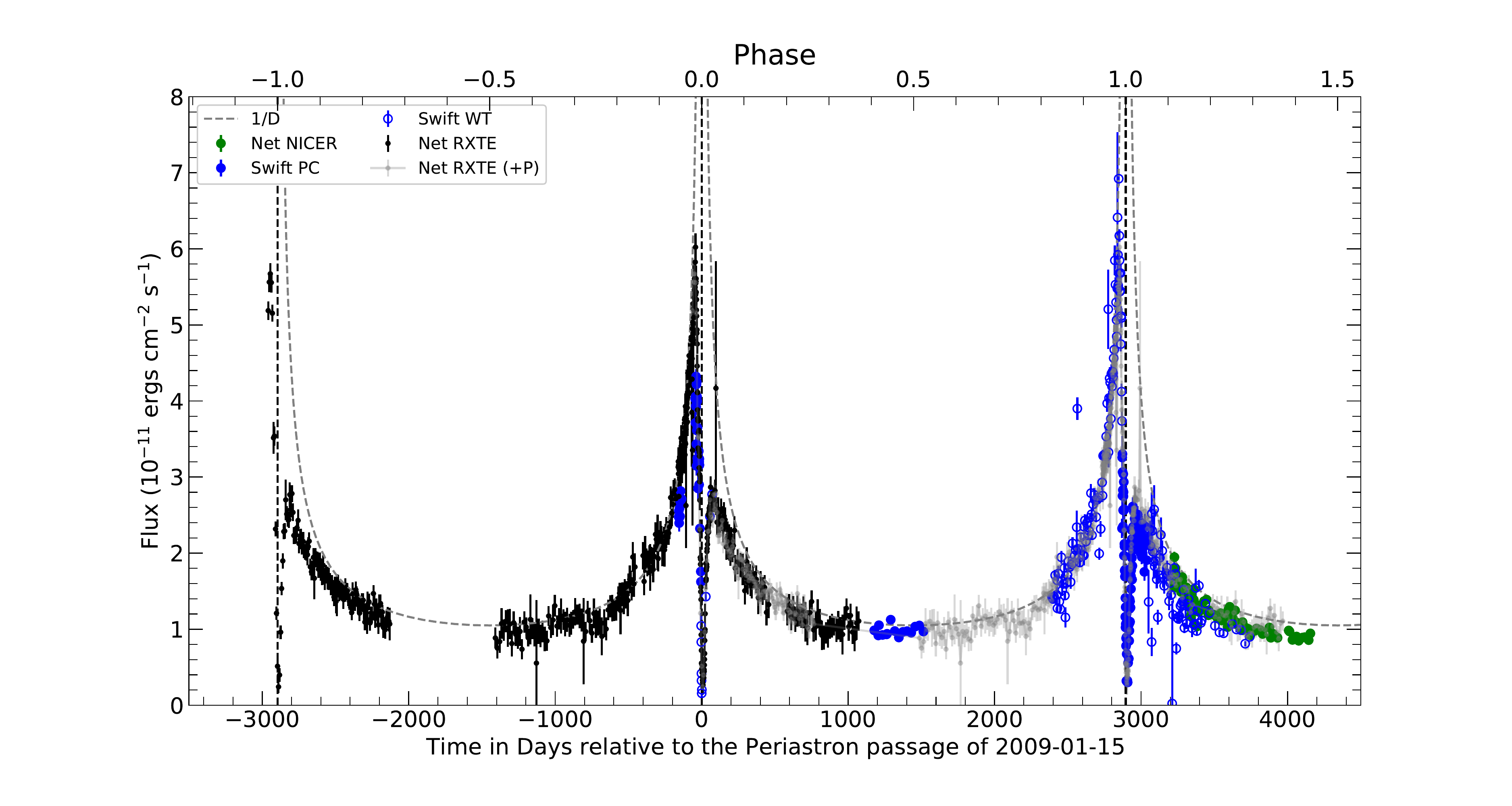}
  \caption{The \rxte, \swift, and \nicer\ lightcurve of \wro, 2000-2020. The time zero point corresponds to the periastron passage %
  JD 2454846.727 =   2009-01-15 05:26.
  The gray points are the RXTE fluxes advanced  by one period. The black dashed vertical lines are times of periastron passages, while the gray dashed curves show the expected $1/D$ variation in flux
  \rev{for an unobscured adiabatic system of colliding winds}. 
 }
   \label{fig:swift_rxte_nicer}
\end{sidewaysfigure}

\begin{figure}[htb!] %
   \centering
    \includegraphics[width=7.5in]{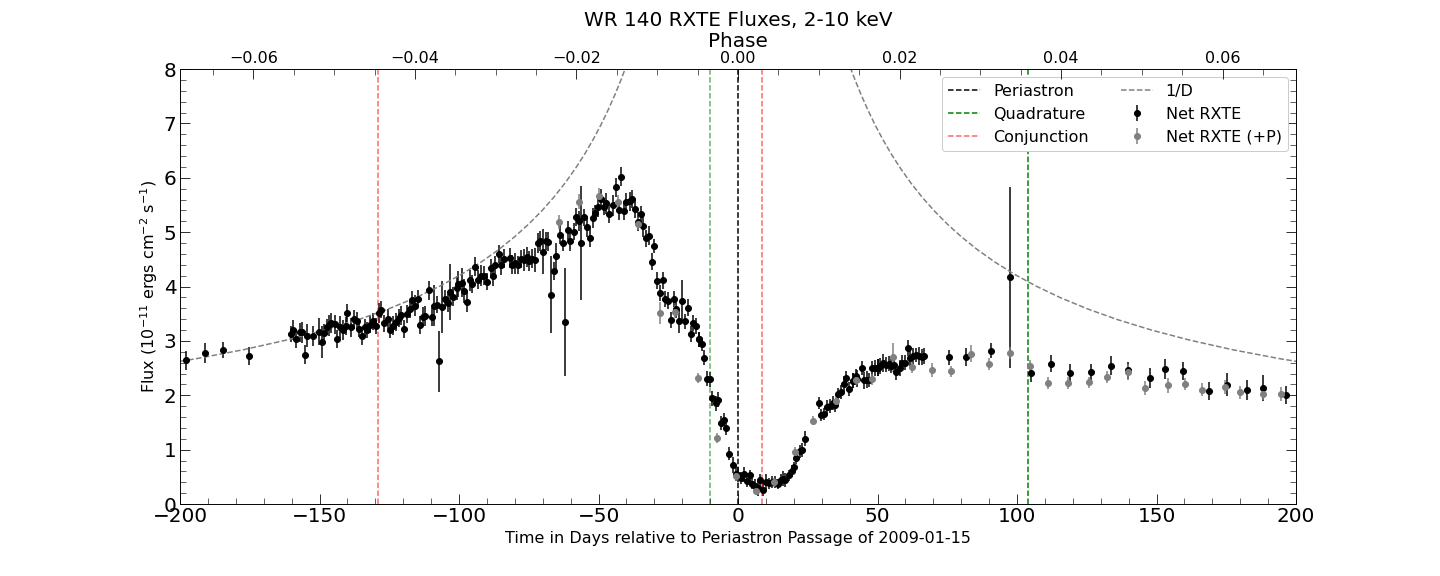}
  \caption{Comparison of the two minima observed by \rxte\ in 2001 (gray circles) and 2009 (black circles). Also shown are the timings of quadratures and conjunctions, with periastron passage in black. WC-star inferior conjunction occurs very close in time to the minimum of the X-ray flux a significant fraction of which is due to the cosmic X-ray background. The gray dashed curve shows a model flux inversely proportional to the stellar separation ($D$).}
   \label{fig:xte_alone_zoom.png}
\end{figure}

\begin{figure*}[htb!] %
   \centering
    \includegraphics[width=7.5in]{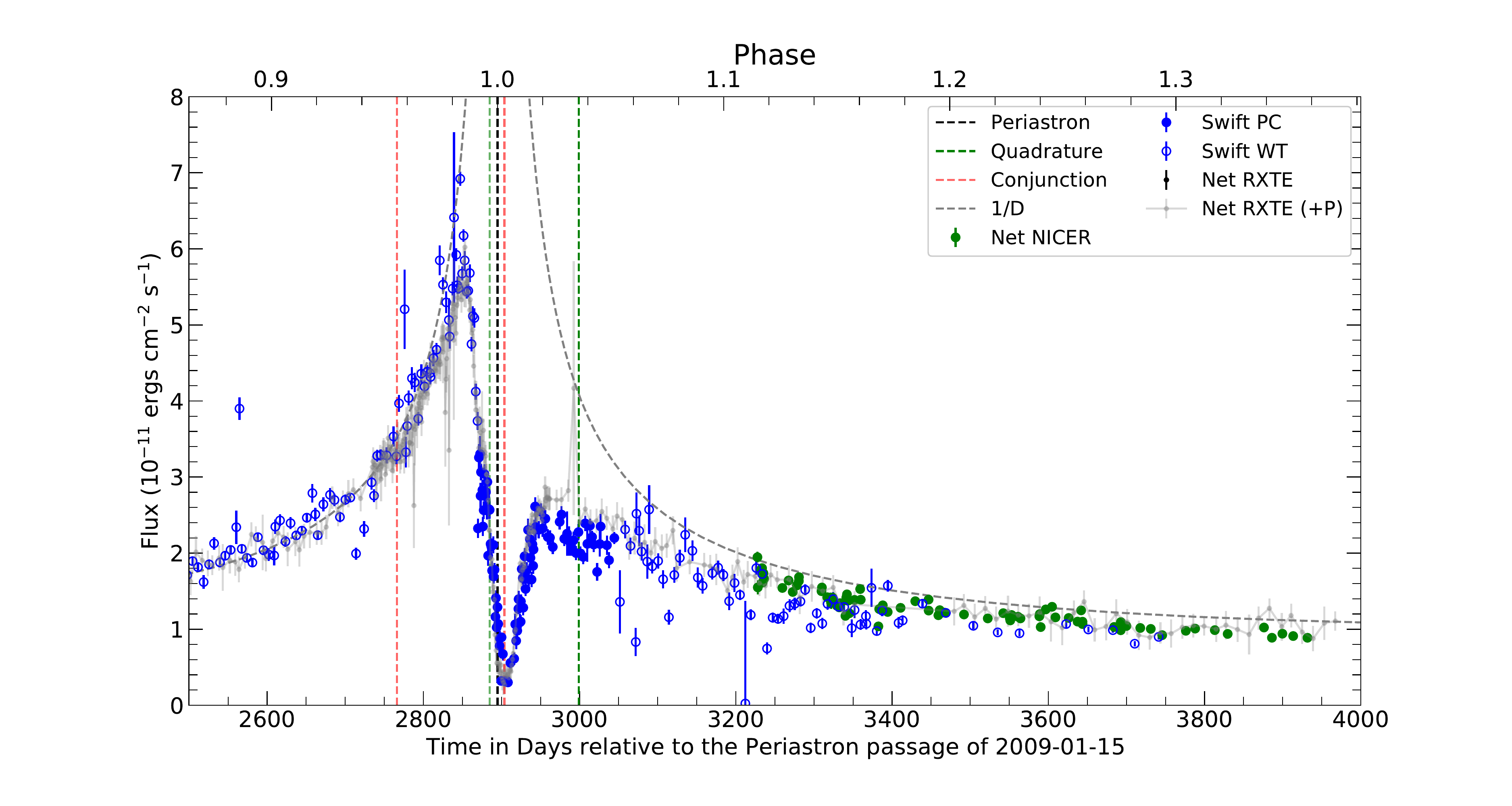}
   \caption{Comparison of \nicer\ and \swift\ contemporanous 2--10 keV X-ray fluxes around the 2016 minimum.  \rxte\ fluxes from the 2009 minimum are also shown.  \swift\ PC-mode data 
are shown as filled blue circles, while WT-mode data are shown as open blue circles.   }
   \label{fig:swift_nicer}
\end{figure*}

Figure \ref{fig:swift_rxte_nicer} shows the complete \rxte, \swift, and \nicer\ 2--10 keV band X-ray flux curve of \wro\ %
covering  {three}~periastron passages.  The general 
characteristic of \wro's X-ray flux curve is a gradual increase in X-ray brightness as the stars approach periastron, followed by a rapid decline to a brief, sharp minimum, a quick recovery from the minimum, and a gradual decline in brightness up to apastron. While this is similar to the X-ray variation seen in \wro's ``sister star'', 
\ec\footnote{An interesting numerological coincidence between these sister binaries: in 2009 \wro\ reached its X-ray minimum on 24 January 2009, just 8 days after \ec\ began its own X-ray ``deep minimum'' phase (on 16 January 2009) near its periastron passage. This coincidence will not happen again for another 400 years.  The next time the X-ray minima will be similarly coincident will be 
in 2413 CE, when \ec's minimum will occur on 29 July, just three days before \wro's own X-ray minimum on 1 August.} \label{sec:eccmp}, another long-period, highly eccentric colliding wind binary, it is worth noting that, unlike \ec, there are no strong, rapid ($\sim$weeks) X-ray variations  \citep[``flaring'',][]{2009ApJ...707..693M} of the X-ray brightness seen in \wro\ just prior to the start of its X-ray minimum. As shown in Figures \ref{fig:swift_rxte_nicer} and  \ref{fig:xte_alone_zoom.png}, the X-ray fluxes from the two orbits observed by \rxte\ agree very well with each other at all common phases. In addition the depth and duration of \wro's X-ray minima are nearly constant from one orbital cycle to another. This can also be contrasted with \ec\ which shows significant changes in the duration of X-ray minimum from cycle to cycle  \citep{2010ApJ...725.1528C, 2017ApJ...838...45C}.

If the cooling of the shocked gas is effectively adiabatic and wind absorption is minimal, the luminosity of X-ray flux with phase is expected to vary inversely with  $D$, the separation between the two stars \citep{usov92,Stevens:1992yu}.   In the figures the gray dashed curve represents a $1/D$ variation and it is clear that near X-ray maximum \& minimum the X-ray brightness curve does not follow  the expected $1/D$ variation.
As can be seen more clearly in Figure \ref{fig:xte_alone_zoom.png}, the deviation from the $1/D$ dependence in the X-ray flux starts about 120 days prior to periastron passage, i.e. around the time of superior conjunction of the WC star.  

Figure \ref{fig:swift_nicer} shows the \swift\ X-ray fluxes in the 2--10 keV band around the 2016 minimum, along with \nicer\ 2--10 keV band fluxes. \nicer\ began observations in July 2017, so that observations started after periastron passage ($\phi= 2.1$). There is good agreement between the \swift\ and \nicer\ contemporaneous fluxes.  As shown in Figure~\ref{fig:swift_rxte_nicer} and in Figure~\ref{fig:swift_nicer}, the \nicer\ fluxes seem to be about 10\% below the projected level of the \rxte\ fluxes at these phases (but in good agreement with the \swift\ fluxes in the area of overlap).

The exceptional X-ray brightness of \wro\ enables accurate assessment of its variability. Even the first few isolated observations of \wro\ with \exosat, \rosat, and \asca\  
 showed that variability was an obvious feature of its behaviour in line with broad expectations of the eccentric binary colliding-wind model. The detailed monitoring built up since over the last two decades through nearly a thousand separate observations with \rxte, \swift, and \nicer\ allows a fairly complete picture to be drawn of the phase-dependent balance between downstream shock emission from the interacting winds and upstream foreground absorption during transfer principally through the more powerful wind of the WC star. As the ensemble of data under consideration here shows no evidence of changes in the equivalent temperature structure of the emission, discussion may properly be framed in terms of the independent measures of luminosity and absorption of the X-ray emission with reference where necessary to where this simple scheme fails, notably at low energies near periastron.
 
The X-ray observations also show that the X-ray spectrum of \wro\  is repeatable at all phases between orbital cycles in both emission and absorption and varies smoothly and monotonically between
\rev{a} 
few turning points that emerge due to boundary conditions of macroscopic orbital dynamics or microscopic X-ray emission physics. Of particular note is the absence of flares or other short-term episodes. Instances of apparent rapid variability between individual exposures visible in some of the figures are invariably due to instrumental instabilities. These ideas will be subject to stricter test in work underway on the ensemble of the few dozen X-ray observations of \wro\ of much higher statistical weight at scattered phases made with other instruments.
 
\subsection{Orbital Dependence of \wro's X-ray Luminosity}

\begin{figure}[htbp] %
   \centering
    \includegraphics[width=7in]{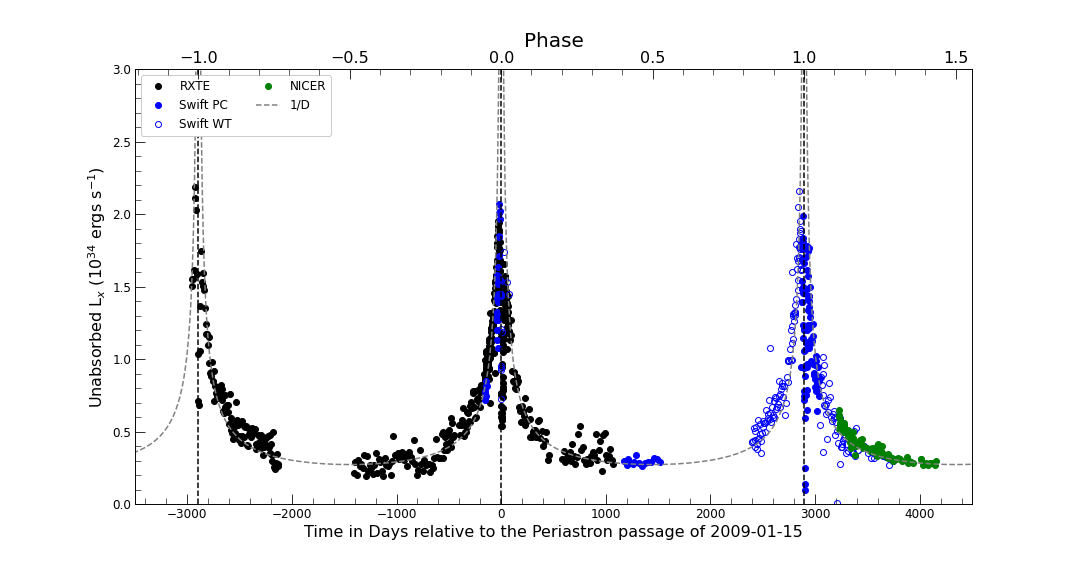} 
  \caption{X-ray luminosity in the 2--10 keV band corrected for interstellar and
  circumstellar absorption assuming the distance in Table \ref{tab:params}.  The dashed line shows a $1/D$ variation.}
   \label{fig:Lxunabs}
\end{figure}

\begin{figure}[htbp] %
   \centering
    \includegraphics[width=7in]{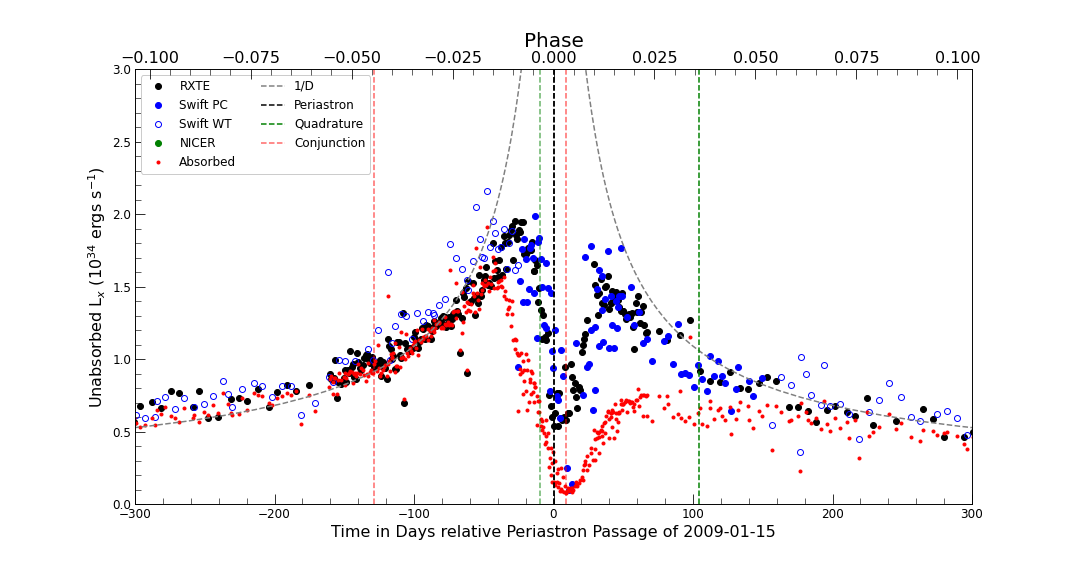} 
  \caption{X-ray luminosity in the 2--10 keV band  {corrected for absorption }in the interval 300 days around periastron passage.  The red points show the observed (i.e. absorbed) luminosity from the \rxte\ and \swift\ observations.}
   \label{fig:Lxunabsperi}
\end{figure}

As shown chronologically for the different instruments in Figure~\ref{fig:Lxunabs} and in folded form near periastron in Figure~\ref{fig:Lxunabsperi}, the intrinsic X-ray luminosity of \wro\ evolves smoothly as a function of binary phase.
At apastron, when the binary system is at its maximum separation, the X-ray luminosity is a minimum. During the subsequent long, nearly 4-year approach to periastron, covered in its entirety from 2005 March by \rxte, the luminosity gradually increases in inverse proportion to the binary separation until about 6 months before periastron when it has doubled, behaviour confirmed in its later stages by the long-term \swift\ campaign initiated 16 months before the next periastron passage in 2016 December. After this point, the luminosity continues its steadily accelerating increase, though now slower than expected from the inverse binary separation, until reaching a maximum, not at periastron, but about 20 days beforehand after which a rapid decrease occurs by about a factor of five over about 20 days. This unexpected decrease corresponds closely, as shown below, with the complementary increase of excess emission in the \CIII~$\lambda$5696 optical emission line in \wro\ \citep{Fahed:2011qy}.

Spectral properties %
close to the X-ray minimum have been relatively hard to establish for a combination of reasons: the observed count rates are lowered even further than just the decreasing luminosity
by heavy and changing absorption during the transit of the WC star; the unresolved \rxte\ background becomes much more significant; and the appearance of the relic component. Despite these challenges, it is clear the minimum is relatively flat for 10-15 days near periastron and conjunction before recovery gets underway. Furthermore, in common with the rest of the X-ray orbit, there is little evidence of the type of secular changes seen during repeated observations of \ec's X-ray minimum.

Despite fewer observations and the higher absorption described below, the recovery after periastron generally mirrors its previous evolution with some asymmetry as the luminosity appears to reach its local maximum perhaps 10 days later than expected from its counterpart before periastron. A few weeks later, symmetry in luminosity seems to have been established with good agreement between the three instruments during a steady decrease towards apastron inversely proportional to the binary separation.

 \subsection{Orbital dependence of \wro's X-ray absorption}
\label{sec:nh}

\begin{figure}[htbp] %
   \centering
 \includegraphics[width=7in]{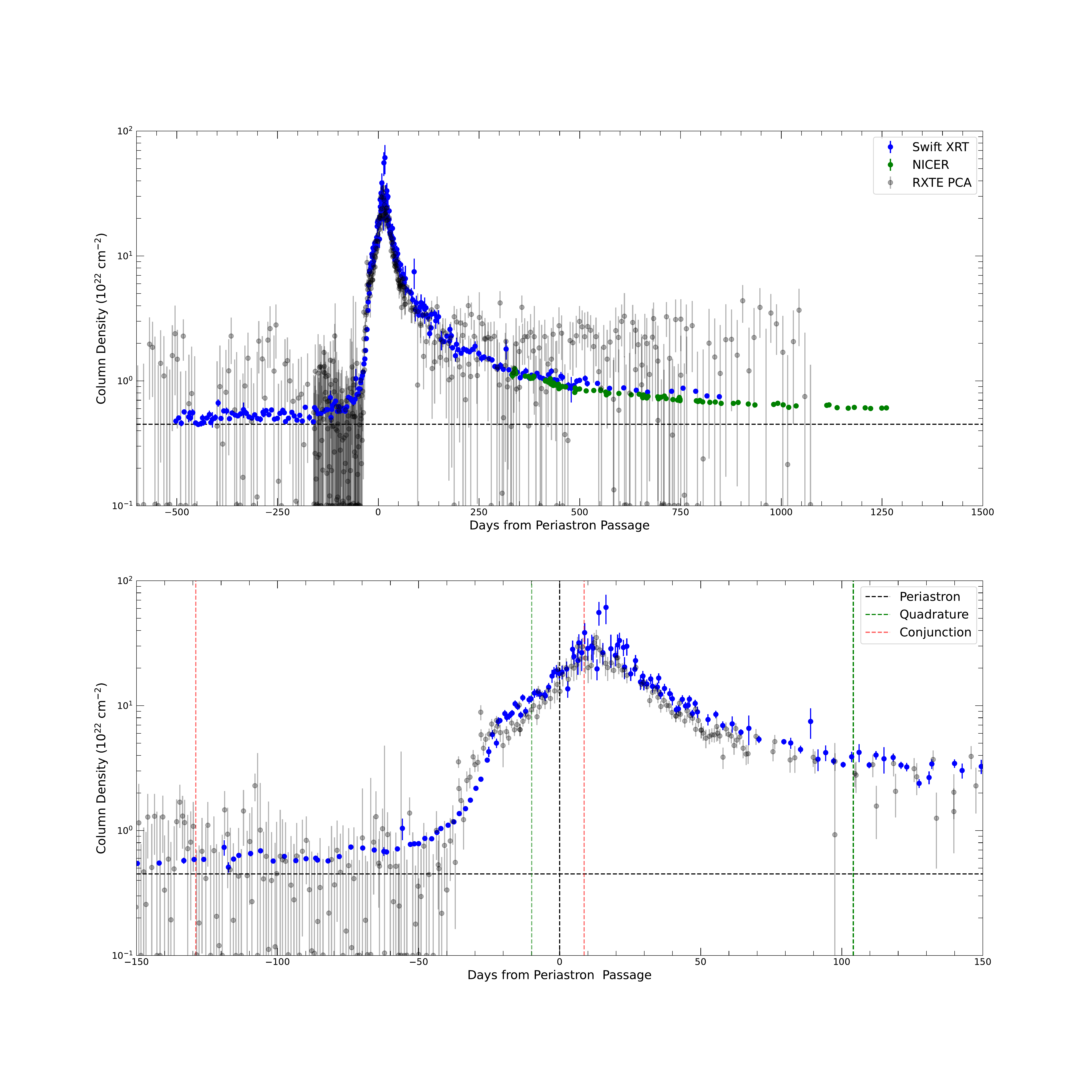} 
  \caption{\textit{Top}: Column density variations from best fits to \rxte, \swift\ and \nicer\ spectra. { A horizontal dashed blue line 
shows the column density far from periastron}, i.e. $N_{H}= 4.6\times10^{21}$~cm$^{-2}$, which represents the interstellar column to the system without additional wind absorption. \textit{Bottom}: Same as above emphasizing the 300 day interval around periastron passage. Red, green and black  {vertical} lines mark times of conjunctions, quadratures and periastron passage, respectively. }
   \label{fig:nh}
\end{figure}

As shown in Figure~\ref{fig:nh} on a logarithmic column-density scale, the variable X-ray absorption observed in \wro\ has a complex structure of high dynamic range that is asymmetric with orbital phase.
Agreement is good between measurements by \rxte\ and \swift\ near the periastron passages of 2009 and 2016, indicating that the wind absorption stayed repeatable with orbital phase over at least the last two orbital cycles. \nicer\ and \swift\ also agree well at common phases.

The variations in column density are consistent with general expectations of the colliding-wind shock model in which the upstream O-star wind is confined by the more powerful WC-star wind within a cone-shaped boundary having a half-opening angle relative to the line of centers of $\theta_S\approx32\degree$ -- 41\degree according to the rough radiative and adiabatic estimates in Table~\ref{tab:params}. The X-ray emission occurs in the intervening interaction region that forms an extensive interface between the two upstream winds of dimension comparable with the binary separation. The carbon-rich WC-star wind is responsible for most of the absorption around the orbit, shown by the maximum within a few days of WC-star {inferior} conjunction when the WC-star is in front of the system's center-of-mass,  the stagnation point and its stellar companion. The high eccentricity and the longitude of periastron $\omega$ determine that the WC star is nearer the observer for only about 138 days including periastron, while the O star is in front for the remaining 2757 days, including apastron.

Significant absorption path lengths through the wind of the O star occur near its inferior conjunction for orbital inclinations $\sin{i}>\cos{\theta_S}$, i.e.,
 $i > 58^{\circ}$. 
 For \wro, although Table~\ref{tab:params} suggests that $\sin{i}\approx\cos{\theta_S}$, the obvious transition in the column density measurements over the short interval between the two conjunctions shows that the winds of both stars are involved, allowing among other things estimates to be made below of the system geometry. Other slower variations with orbital phase are partly due to changing absorption path lengths through the two winds.

A basis of the hydrodynamic models often used for simulations of colliding-wind systems is the absence of any information to the flow of upstream gas that would signal the presence of shocks further downstream. In this case, foreground absorption is identical to that of single stars, allowing the construction of simple geometrical models. The column density integrated along a path entirely through the terminal velocity flow of a wind from a point $\vec{R}_{x}$ to a distant observer takes the concise form
\begin{equation}
N_{x}~\mathrm{cm}^{-2} = 4.31\times10^{23} \frac{\dot{M}_{-6}}{\mu v_8} \frac{\Theta}{b} + N_{\mathrm{ISM}}
\label{eq:Nx}
\end{equation}
where $\dot{M}_{-6}\times10^{-6}$~\msun~yr$^{-1}$ is the mass-loss rate;
$\mu$ is the mean atomic weight;
$v_8\times1000$~\kms\ is the wind terminal velocity;
$\Theta$~radians is the aspect angle between the vector joining the stars and the line-of-sight; 
$b = R_x\sin{\Theta}$~\rsun\ is the impact parameter;
and $N_{\mathrm{ISM}}$ is the constant interstellar term.
In \wro, the geometrical terms are accurately defined by the well-constrained orbit.

\begin{figure}[htbp] %
   \centering
   \includegraphics[width=6in]{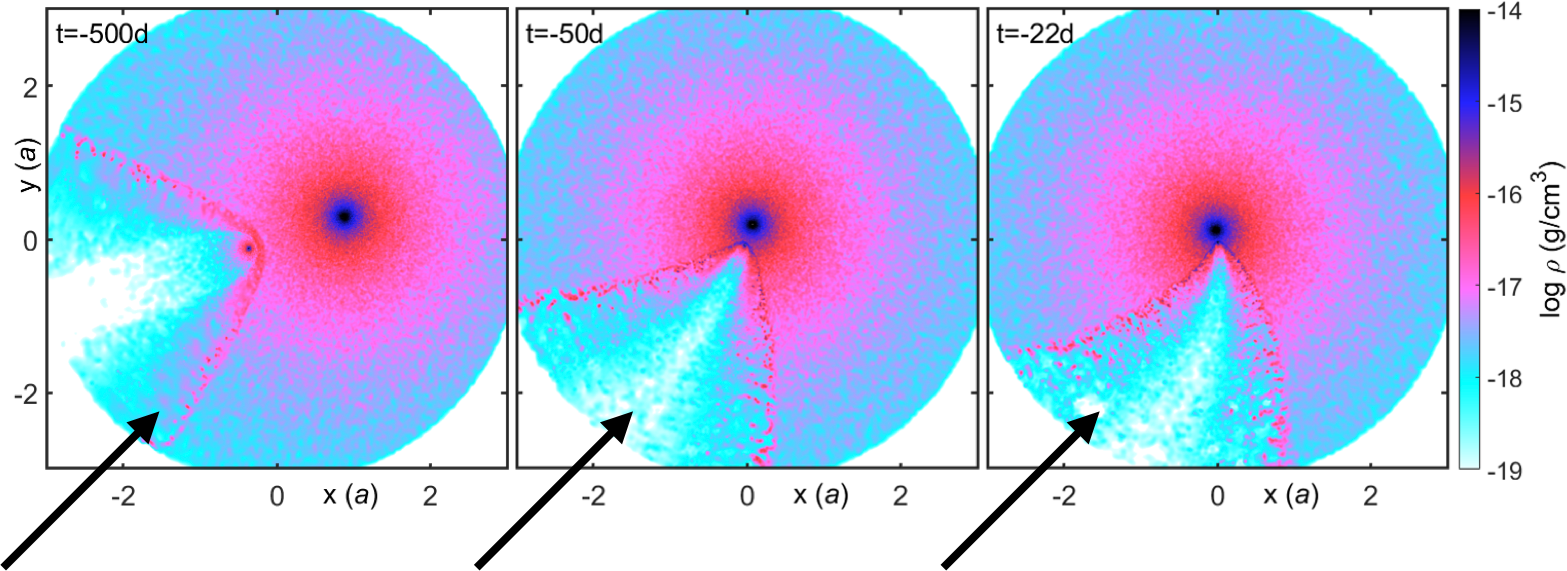} 
   \caption{Simulated density distributions of \wro\ in the orbital plane at $500$~days (left), $50$~days (middle) and $22$ days before periastron passage from a dynamic 3-D smoothed-particle hydrodynamics fluid model (2021, C. Russell et al., in prep.).  The arrow represents the observer's line-of-sight projected onto the orbital plane with the spatial scale in units of the semi-major axis, $a$. The half-opening angle of the shock cone in this simulation is about $40\degree$. At $-500$ days the leading edge of the shock cone around the O star has recently crossed the observer's line-of-sight.  The shock cone takes about $450$ days to pass the line-of-sight.  Between $-50$ and $-22$ days, the trailing edge of the shock cone passes the line-of-sight which shortly afterwards encounters the dense unshocked wind of the WC7 star, causing a rapid increase in column density as the stagnation point of the shock passes close to and behind the WC7 star.}
   \label{fig:cwbsim}
\end{figure}

Although the X-ray source is likely to be comparable in size to the binary separation, for simplicity we have calculated indicative column densities along single lines-of-sight from the stagnation point as a function of phase. The single-point approximation provided good fits to all the data considered here and has been widely used in the literature for the analysis of colliding-wind systems. Its use is justified by examples that we have calculated of X-ray sources extended along the analytical shell interface of \citet{1996ApJ...469..729C}. More sophisticated approximations are likely to require detailed dynamical simulations. For example, in order to anticipate the type of work that will be required and to illustrate for present purposes the system geometry at significant phases in the column density curve from the beginning of the \swift\ campaign, Figure \ref{fig:cwbsim} shows density distributions in the orbital plane of \wro\ from 3-D smoothed-particle hydrodynamics simulations \citep{Russell:2013qf} based on the parameters in Table \ref{tab:params}.

\subsubsection{Absorption by the O-star wind}
\label{section:NxO}

\begin{figure}[htbp] %
   \centering
   \includegraphics[width=6in]{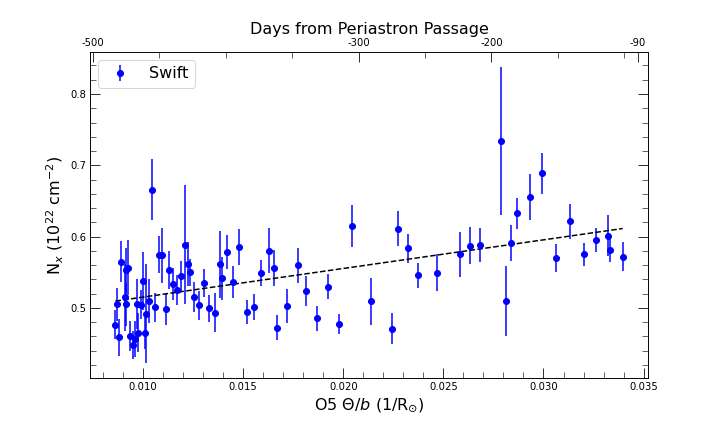} 
   \caption{X-ray absorption column density observed in \wro\ with the \swift~XRT as a function of the column-density integral through the O-star wind given by ratio of the aspect angle ($\Theta$) to impact parameter $b$ in the range 500-80 days before periastron passage.  The dashed line is the linear fit to the data,
   $(3.956\times\Theta/b+0.464)\times10^{22}$~cm$^{-2}$.}
   \label{fig:O5Nx}
\end{figure}

While maximum absorption close to WC-star conjunction was already evident from the earliest observations with \rxte, the orbital dependence at the lowest columns has only been revealed with the soft X-ray capabilities of the \swift~XRT campaign that began in 2015 August, about a year and a half before the most recent periastron. In the observations from the beginning of the campaign in the approach to O-star conjunction 129 days before periastron and beyond, the observed column increased slowly but steadily by about 20\% over the course of a year or more until 50 or 60 days before periastron, when a rapid increase in column began. As shown in Figure~\ref{fig:O5Nx}, when the shocks around the O star are directed towards the observer and the WC7-star wind makes its smallest contribution, this lowest part of the absorption curve observed so far between 500 and 80 days before periastron is well modelled by a combination of a constant component due to the interstellar medium and a term linear in $\Theta/b$ following Eq.~(\ref{eq:Nx}) for the O star. With this model, the estimate of the constant interstellar term is $(4.64\pm0.07)\times10^{21}$~cm$^{-2}$, consistent with the interstellar column in the direction of \wro\ derived from \wro's colour excess or Ly$\alpha$ absorption. For $\mu_{O5}=1.30$, the linear term yields an O-star mass-loss rate of
$\dot{M}_{O5}=(3.7\pm0.3)\times10^{-7}$~\msun~yr$^{-1}$,
about half the smooth-wind value adopted in Table~\ref{tab:params}, which was the lowest among several alternatives discussed by \citet{2006MNRAS.372..801P}.

\begin{figure}[htbp] %
   \centering
   \includegraphics[width=6in]{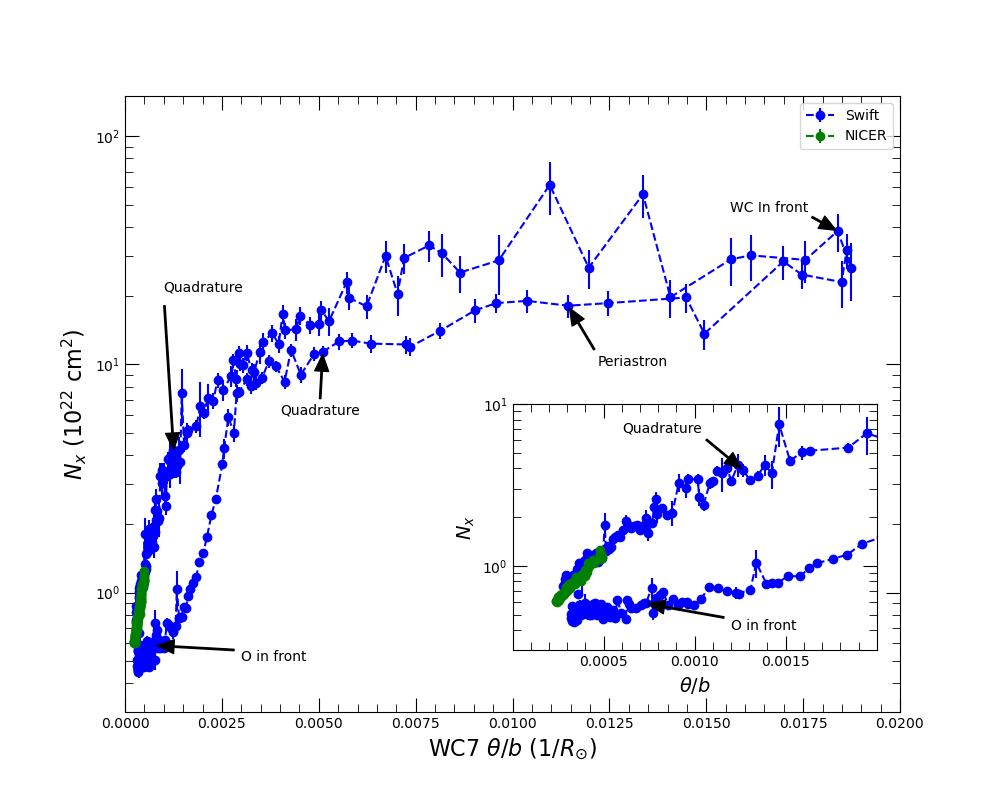} 
   \caption{X-ray absorption column density observed in \wro\ with \swift\ and \nicer\ as a function of the  column-density integral through the WC-star wind at all available phases including those associated with the O star shown in Figure~\ref{fig:O5Nx}.}
   \label{fig:WC7Nx}
\end{figure}

\subsubsection{Transition between O-star and WC-star winds}
\label{section:NxOW}

An interesting and perhaps unexpected feature revealed by both the measurements and the simple model of this highly eccentric binary is that the conjunction of the O star does not coincide with an absorption maximum. Instead, as a result of the rapidly decreasing binary separation, the absorption maximum would have been reached about 100 days later, a month or so before periastron, if other events did not intervene.

The extended episode of low absorption ends between O-star conjunction and quadrature with a rapid increase in column density of more than an order of magnitude in a month as the line-of-sight passes from the O-star wind into the WC-star wind, presumably via an intermediate direct view of shocked gas as the line of sight encounters the wall of the shock cone. The transition begins slowly about 60 days before periastron before increasing rapidly until an inflection point 22 days before periastron when the line-of-sight starts to pass predominantly through the WC-star wind. These phases mark the angular positions with respect to the system axis of the two shock discontinuities corresponding to aspect angles of $34.6\degree$ at the inner O-star shock and $61.2\degree$ at the outer WC-star shock. A contact discontinuity, whose existence might be doubtful in a collisionless plasma, would be located at some intermediate value. In any case, this estimate of the angular extent of the shock is within a few degrees of the theoretical values given in Table \ref{tab:params}. 
\rev{It} 
agrees well with the estimate of $34\degree\pm1\degree$ for the angle of the contact discontinuity inside the shock[s] derived from \ion{He}{1}$~\lambda$10830 absorption \citep{2021MNRAS.503..643W} and is reproduced with reasonable success by the fluid simulations shown in Figure~\ref{fig:cwbsim}. 

\subsubsection{Absorption by the WC-star wind}
\label{section:NxW}

Once the WC-star wind intervenes, absorption becomes much stronger due to the appearance of material enriched in carbon and oxygen and a higher mass-loss rate which combine easily to outweigh the greater distance of the shock from the Wolf-Rayet star. It should be borne in mind that in order to make a fair comparison of absorption at all orbital phases, the same solar abundance photoelectric model has been used throughout and any mass-loss rate estimates need to take this into account. When the 30--40 day transition is over, the column density continues to increase for another 30 days at a slower but more regular pace through quadrature and periastron to reach a maximum close to WC-star inferior conjunction. At this point, for absorbing material of Table~\ref{tab:templ_mo} emission abundances with $\mu_{WC7}=5.73$ and correspondingly higher photoelectric cross-sections, the implied mass-loss rate is
$\dot{M}_{WC7}=1.7\times10^{-5}$~\msun~yr$^{-1}$, a factor of 3
smaller than the value given in Table \ref{tab:params}.

The post-periastron orbital dynamics are slow in comparison so that the aspect angle of $61.2\degree$ marking the WC-star shock encountered 22 days before periastron is reached only 654 days afterwards at much greater orbital separation. The binary dynamics of this highly eccentric system are therefore the major cause of the comparatively slow decrease in absorption over years after WC-star conjunction compared with the rapid evolution over a month before.
However, in contrast to the absorption through the O-star wind, the same type of simple linear geometrical model has no success reproducing the higher, longer-lived WC-star absorption. Instead, the absorption follows the distinctive locus in the $N_x-\Theta/b$ plane shown in Figure~\ref{fig:WC7Nx} that was constructed from all the available column density measurements made with \swift\ and \nicer\ from 502 days before periastron to 1261 days after. This amounts to 61\% of the orbit where the stars are moving at their slowest leading to the expectation that the small unexplored part of the locus at the lowest column densities will take just over 3 years to complete and probably close.

The locus consists of two loops: an upper, rapid, roughly horizontal loop associated with the WC-star wind and high column densities that lasts about 70 days when the stars are moving at their greatest orbital velocities; and a lower, slower, roughly vertical loop associated with the O-star wind and lower column densities that otherwise occupies nearly the whole orbit. The locations of periastron, conjunctions and quadrature are marked. The two loops coincide close to the inflection point that marks the end of the rapidly ascending transition between the winds of the O star and WC star. Once the highest values have built up relatively slowly towards WC-star conjunction, the column density stays roughly constant for nearly three weeks despite a large reduction in $\Theta/b$. This behaviour strongly suggests the presence of a large amount of absorbing material not connected with the line-of-sight through the undisturbed upstream wind
but instead with the shock itself. This matter is discussed further below where we suggest that reflected ions are responsible.
 
\subsubsection{Transition between WC-star and O-star winds}
\label{section:NxWO}

The rapid transition between O-star wind and WC-star wind intervening in the lines-of-sight to the X-ray source observed a few weeks before periastron through the sharp increase in column density over a month must have a counterpart in the opposite direction after periastron in which the lines-of-sight move from WC-star wind to O-star wind. Neglecting the curvature of the shock fronts, it is plausible that this could take place roughly near the corresponding orientations between aspect angles $61.2\degree$ and $34.6\degree$.  If so, it would involve a slowly decreasing column density starting 654 days after periastron and lasting for more than half the entire orbit over years rather than weeks. The observations taken so far are consistent with this idea with the latest \nicer\ measurements showing gradually falling column densities to values still significantly greater than the lowest values seen at the beginning of the \swift\ campaign. This trend should continue through apastron and require precise measurements to follow its progress for the next few years and complete the loop which might include a shallow local minimum a few months after the binary separation starts to decrease again.

\subsection{The ``Relic'' Component}
\label{sec:relic}

Analysis of \suzaku\ X-ray spectra during the X-ray minimum associated with the 2009 periastron passage by \cite{Sugawara:2015eu} uncovered a soft component below about 2~keV, which they suggested was a relic of cooling plasma from the wind collision.  This ``relic'' component was seen in spectra obtained on 2009-01-04 and 2009-01-13, at orbital phases of 
-0.004
and 
0.001, respectively. The relic component
seems similar to residual soft emission revealed in other systems {  at orbital phases when the} colliding-wind emission suffers severe photoelectric absorption: relevant examples include $\gamma^2$~Velorum \citep{2004A&A...419..215H}, the brightest Wolf-Rayet star in the sky and also a WC+O binary system; and \ec\ \citep{2007ApJ...663..522H} where a so-called ``Central Constant Emission'' component emerges at its deep X-ray minimum.

\begin{figure}[htbp] %
   \centering
   \includegraphics[width=6in]{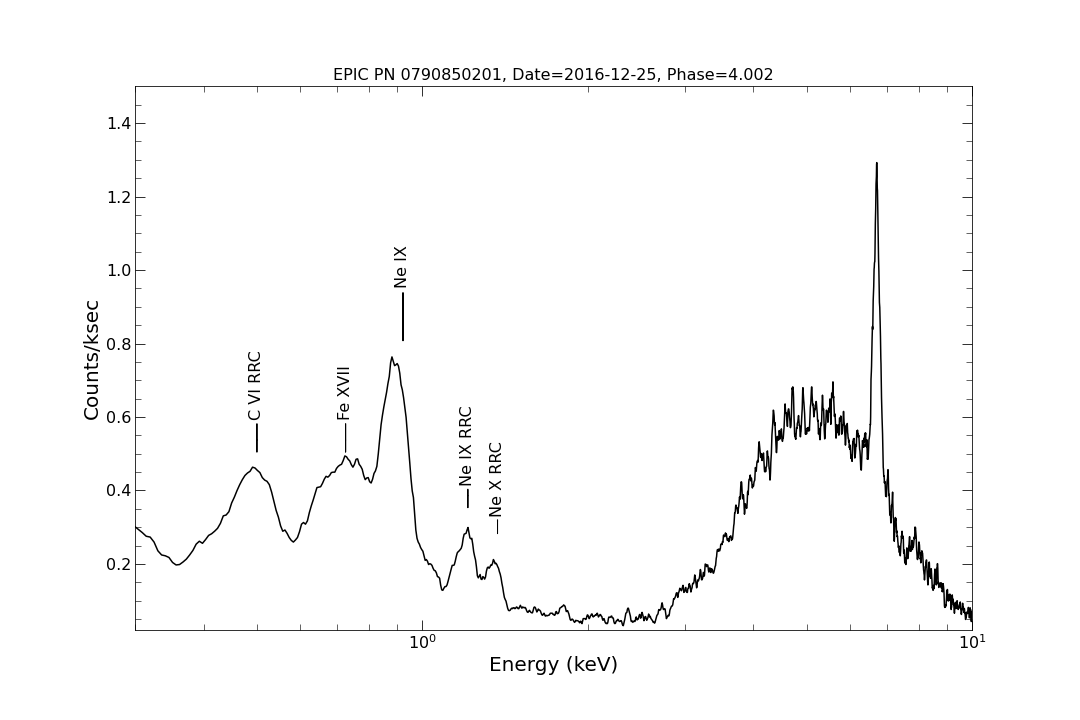} 
  \caption{
\xmm\ EPIC-pn energy spectrum of \wro\ observed about a week after periastron passage in 2016 showing most clearly the relic component emission below 2~keV and suggested identifications for some of its narrow features displayed on a logarithmic scale in energy.  
   }
   \label{fig:relicspec}
\end{figure}

Figure~\ref{fig:relicspec} shows  {a} well-exposed \xmm\ EPIC-pn X-ray spectrum of \wro\ taken two days before maximum X-ray absorption at the inferior conjunction of the WC7 star. It was retrieved from the \xmm\ Science Archive\footnote{\url{http://nxsa.esac.esa.int/nxsa-web/}} under ObsID 0790850201. It shows clear distinction between the hard colliding-wind emission in which the line emission from \ion{Fe}{25} remains prominent above the highly absorbed continuum, and the soft relic component. The relic is composed of narrow features that are resolved even with the limited spectral capabilities of the EPIC-pn instrument and whose energies are well-determined enough to enable the identifications suggested in Figure~\ref{fig:relicspec} following \citet{Sugawara:2015eu}.

\begin{figure}[htbp] %
   \centering
   \includegraphics[width=6in]{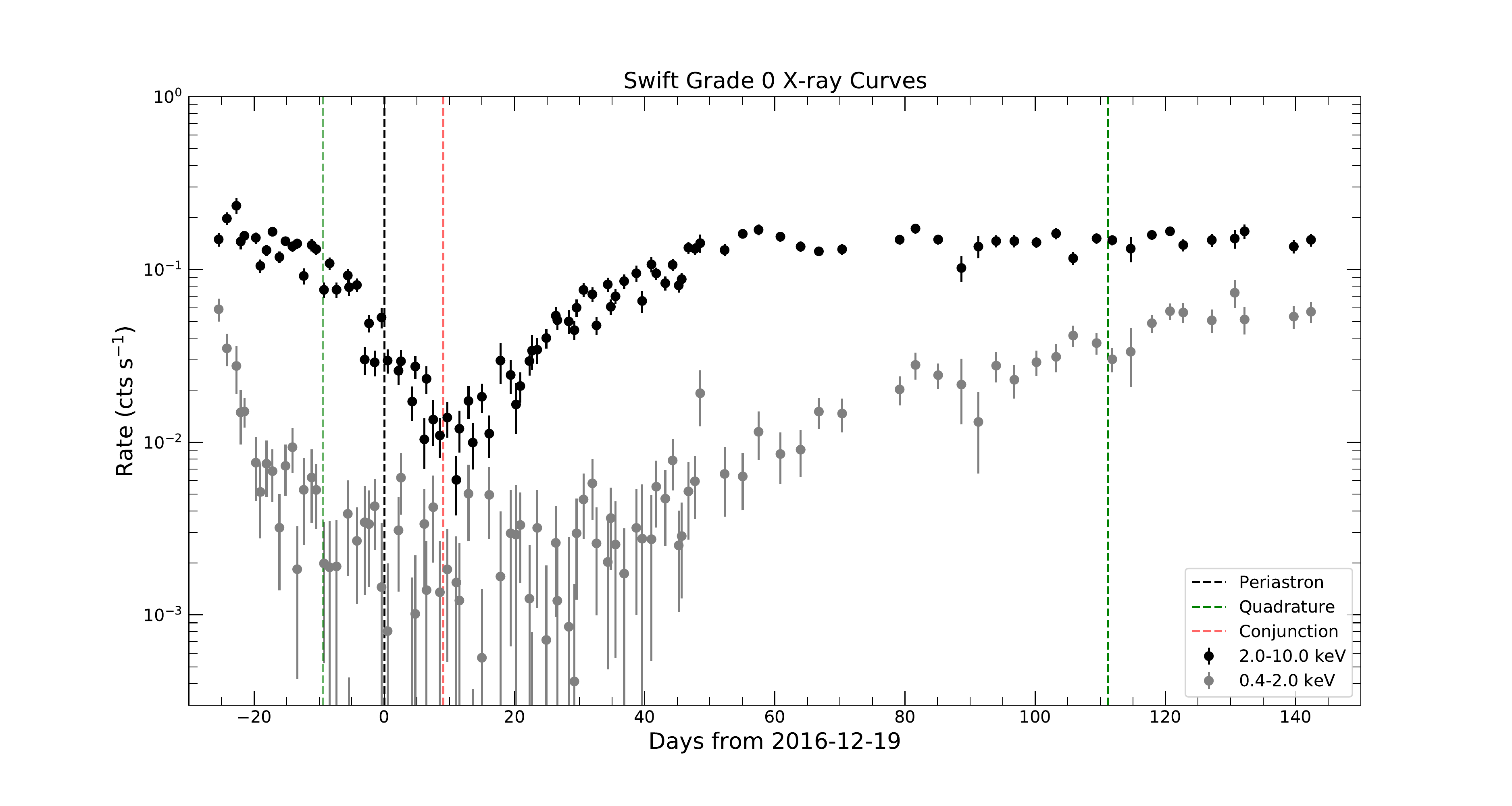} 
  \caption{
The X-ray count rate in the relic band (0.4~--~2.0~keV) from \swift\ XRT observations including only grade 0 events above 0.4 keV, compared to the 2~--~10~keV colliding wind emission from the grade 0 events. The relic component is visible from 25 days before periastron passage to about 50 days afterwards, during times of high absorption. The XMM-Newton spectrum in Figure~\ref{fig:relicspec} was taken close to conjuction
   }
   \label{fig:relic}
\end{figure}

The \swift\ monitoring observations around the 2016 periastron passage provide a measure of the appearance of this component and its temporal behavior once account is taken of the effects of PC-mode optical loading.
The \xmm\ data shown in Figure~\ref{fig:relicspec} do not suffer from this effect.
The XRT was switched to PC mode once the count rate had fallen sufficiently
as \wro\ approached its minimum for pile-up no longer to be significant, since PC-mode provided a directly-determined background estimate, important when the X-ray source was faint.
Initial inspection of the lowest-energy data then showed erratic variability
which proved to be confined to multiple-pixel  (as opposed to single-pixel) X-ray events as classified
by event grade. Event grade is assigned early during data processing to show
the number and pattern of neighboring detector pixels contributing to the recognition
and reconstruction of an individual X-ray photon. Single pixel events are classified as grade 0 while multiples are doubles with grades 1-4; triples with grades 5-8; or quadruples with grades 9-12.
Composite spectra for doubles, triples and quadruples accumulated during the minimum
showed the narrow low-energy peaks at 2, 3 and 4 times the event threshold, respectively, characteristic of optical loading affecting optically bright targets in tailored calibration work. 
Single grade 0 events with energies below 0.4~keV were excluded for the same reason.

Figure \ref{fig:relic} shows the variations of the grade-0 selected count rates in
soft ($0.4<E<2.0$~keV) and hard ($2.0<E<10.0$~keV) bands as a function of phase near periastron passage and WR conjunction a few days later. The soft band includes the relic component while the hard band is dominated by the thermal emission from the colliding wind shock.  The relic component is revealed about 10~--~15~days before periastron due to combination of increasing absorption and decreasing luminosity of the colliding-wind shocks, and is visible for about 50~days  {after periastron}
until absorption and luminosity have recovered for low-energy events
again to be dominated by emission from  {the wind-wind collision}.
During the 50-d minimum, although the relic component is responsible for only a handful of counts in each observation, there is some statistical evidence for variability. The accumulated spectrum closely resembles that seen at occasional relevant phases during much longer observations with \chandra, \xmm, and \suzaku.

\section{The \CIII~$\lambda5696$ Connection}
\label{sec:ciii}

As the stars in \wro\ approach and pass periastron, significant excess emission appears superimposed on the strong flat-topped 
\CIII~$\lambda$5696 emission line 
characteristic of WC stars \citep{1995IAUS..163..460H, 2003ApJ...596.1295M, Fahed:2011qy}. 
The excess has complex and well-resolved velocity structure that moves rapidly from blue to red across the line as the intensity strengthens and fades through periastron and beyond.
Similar excesses are also seen in \CIII~$\lambda5696$ in
other WC colliding-wind binaries \citep[e.g.,][]{2018MNRAS.474.2987H}
and in other lines in \wro\ such as \ion{He}{1}$~\lambda$10830 \citep{2021MNRAS.503..643W}.
It has been natural to suppose such optical and infrared excesses to originate some way downstream in the wind collision zone after shocked WC-star plasma has had time to cool to an appropriate temperature ($T\lesssim 10^{5}$~K) with the velocity profile formed in the continuously changing orientation of the shocked flow wrapping around the O star as the system revolves in its orbit.

However, three observational arguments suggest instead a closer connection between X-rays and \CIII~$\lambda5696$ concerning, in turn, their correlated temporal behavior; the comparative luminosities; and the observed velocity profiles; all considered in the context of the succession of orbital events of quadrature, periastron and conjunction that occur so close together in \wro's highly eccentric but accurately-defined orbit when shocked gas flows away on time scales $D/v_\infty \sim 1$ day or less. During this crucial part of the orbit, many things are changing rapidly at once on a daily basis with the sharp fall in
\rev{intrinsic}
X-ray luminosity and sharp rise in absorption, combining to produce a steeply declining X-ray count rate as the strength of the \CIII~$\lambda5696$ line increases before turning points throw the system into reverse. 

\subsection{Anti-correlated variability of X-ray and \CIII~$\lambda5696$}

\begin{figure}[htbp]
   \centering
   \includegraphics[width=6in]{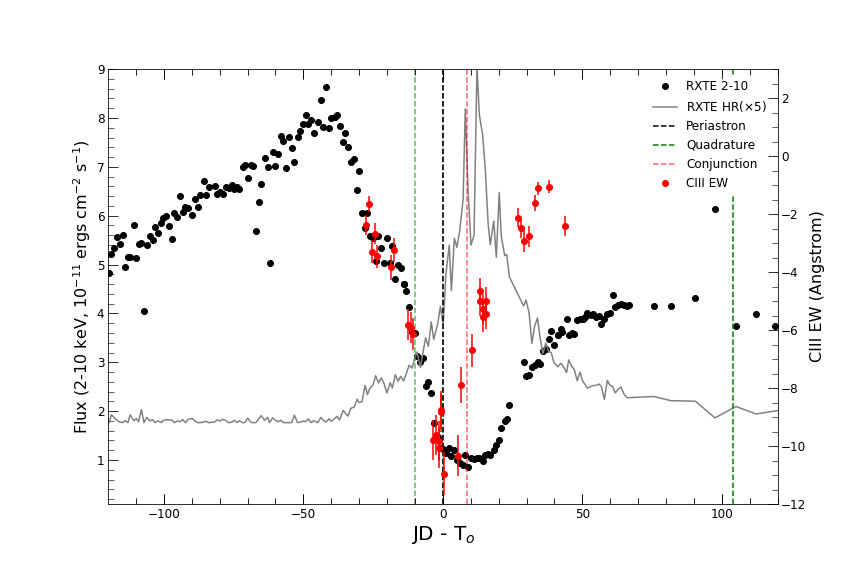} 
   \caption{Comparison of the equivalent width of the \CIII~$\lambda5696$ excess and the X-ray flux seen by \rxte\ during the 2009 periastron passage of \wro.
   The asymmetry in the X-ray curve is mostly due to \rev{photoelectric absorption by the WC-star wind traced by the \rxte\ hardness ratio, HR.}}
   \label{fig:ciii}
\end{figure}

Figure \ref{fig:ciii} compares the evolution of the equivalent widths of \CIII~$\lambda5696$ during the 2009 periastron passage \citep[from Table~7 in][]{Fahed:2011qy} with the 2--10 keV X-ray flux measured during that time by \rxte.  This shows an exceptionally tight correlation between the observed drop in X-ray flux and the strengthening \CIII\ before periastron: formally, the Pearson correlation coefficient = $-0.963$ in the interval between 2008-12-18 and 2009-01-21. About 20 days before periastron, a step occurs in the X-ray decline mimicking closely a similarly subtly-shaped feature in the strengthening \CIII\ line. The lack of any apparent lag between the two morphologies coupled with the timing of the turning points of \CIII\ maximum and X-ray minimum close to periastron suggests a common central origin of both types of radiation. 

\subsection{Luminosities of X-rays and \CIII~$\lambda5696$}

\begin{figure}[htbp] %
   \centering
   \includegraphics[width=6in]{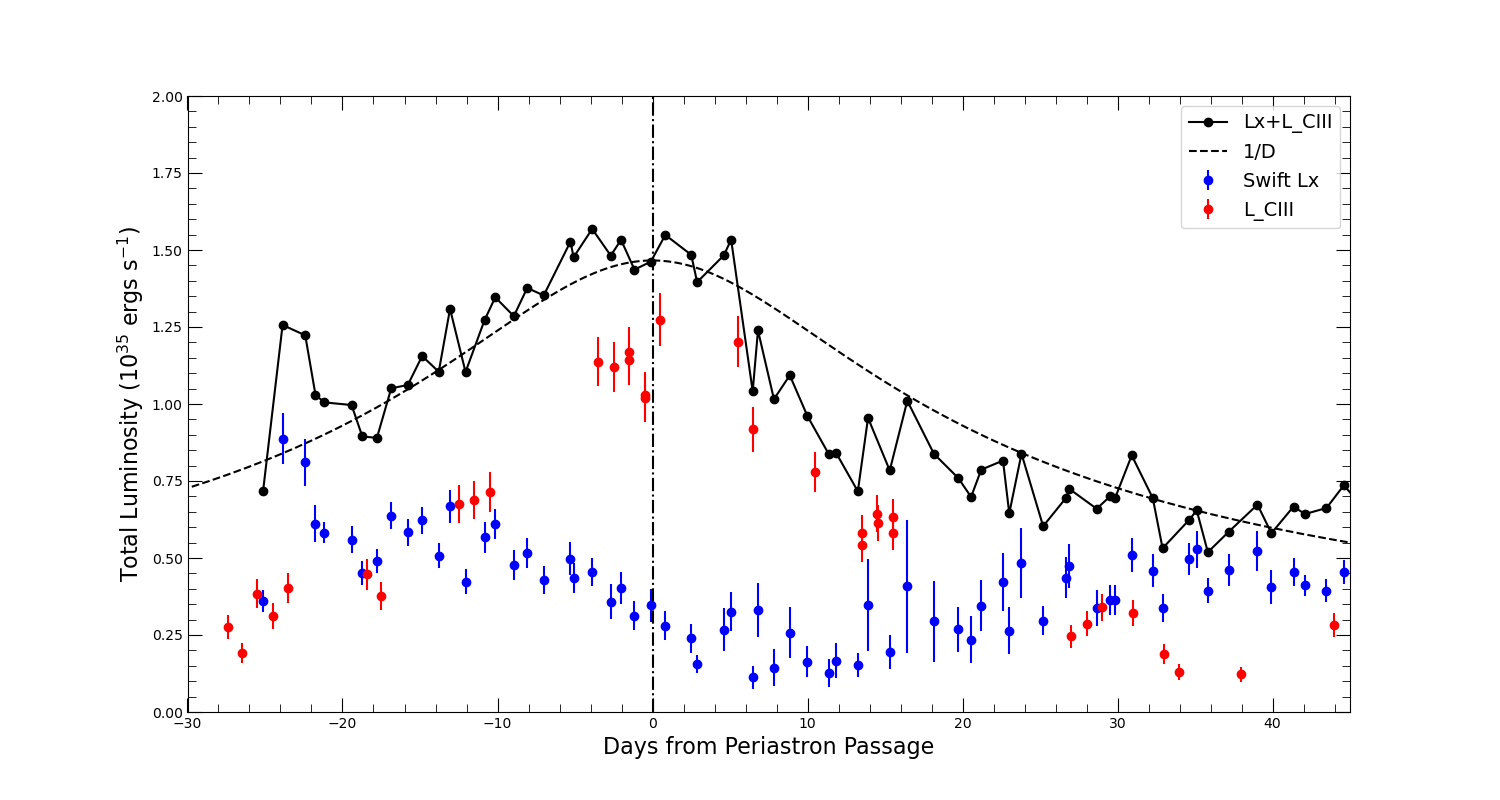} 
   \caption{Comparison of the 0.5--10.0 X-ray luminosity and the luminosity of the \CIII line excess, both corrected for absorption, and the combined (Lx + \CIII) luminosity. The dashed line is a $1/D$ variation scaled by eye to the data.}
   \label{fig:LTOT}
\end{figure}

We converted  the \CIII~$\lambda5696$ excess equivalent widths reported by \citet{Fahed:2011qy} to luminosities to compare to the phase-dependent intrinsic X-ray luminosities and assess the relative importance of the X-ray and \CIII\  in the overall shock energy budget during the few weeks around periastron when the excess appears. 
For \wro\ values of $V=6.85$ and $A_{V}=2.01$ and the standard calibration of the $V$-magnitude intensity scale of
$f_{\lambda}= 3.64\times10^{-9}$~erg~cm$^{-2}$~s$^{-1}$~\AA$^{-1}$ at $V=0$ gives a conversion factor of 
1.16$\times10^{34}$~erg~s$^{-1}$~\AA$^{-1}$ for a distance of 1.52~kpc to \wro.  Figure~\ref{fig:LTOT} shows immediately the close complementary relationship between the X-ray and \CIII\ luminosities in which the central X-ray deficit seen in Figure~\ref{fig:Lxunabs} is almost exactly compensated by \CIII. The combined luminosity $L_{X+C III}=L_{X}+L_{C III}$ formed by these two components of the energy budget and derived by interpolation is also shown and closely resembles the ideal $1/D$ prediction
of colliding-wind theory. The combined luminosity observed at periastron, where the \ion{C}{3}~$\lambda5696$ excess peaks, is $L_{X+\CIII}=1.8\pm0.1\times10^{35}~$\lumcgs and the X-ray luminosity observed at apastron, in the expected absence of a \CIII\ excess, is $L_{X}=1\times10^{34}$~\lumcgs. Their ratio, $18\pm1$, is consistent with the value of $(1+e)/(1-e)=18.9$, where $e$ is the orbital eccentricity, expected if the $1/D$ prediction applies around the entire orbit of \wro. This is encouraging, though it should be borne in mind that the error given in the ratio, determined by the reported uncertainty in the \CIII\ equivalent width, is probably an underestimate due to unknown systematic errors that apply, for example, to the reddening correction.

\subsection{Velocity structure of the \CIII\ excess}

Information about the dynamics and location of the material radiating in the \CIII~$\lambda5696$ line is available by considering, in addition to its intensity, the variable velocity profiles of the excess shown in Figure 3 and plotted in Figure 5 of \citet{Fahed:2011qy}. The observations illustrated there fall conveniently into separate groups according to orbital geometry. Before quadrature, when the WC star is behind its O-star companion, the excess, though faint in its early development, is nearly all confined as expected to negative velocities as the radiating material flows towards the observer.  A week later, in a set of 3 observations more-or-less at quadrature, the excess is distributed roughly evenly between blue and red velocities. The maximum observed red shift is reached within perhaps 2 or 3 days of conjunction when the WC star is directly in front of its companion and shocked material is flowing away.

\section{Discussion}
\label{sec:discuss}

\rxte, \swift, and \nicer\ X-ray monitoring observations of \wro\ have confirmed that its X-ray properties varied in a predictable way over the 4 orbital cycles between 2000 and 2020.  They show that the shock conditions and radiation transfer vary repeatably as a function of orbital phase and therefore provide a rich and detailed set of measurements on which improved understanding of the underlying physics may be built. Appropriate 3D models combining orbital and gas dynamics with the physics of shock formation and dissipation are likely to be a challenging enterprise but whose boundary conditions are now more clearly defined by the results presented here concerning the cooling mechanisms of collisionless shocks under different conditions and the geometry of the interacting winds.

With its uniquely well-determined 3D highly-eccentric orbit, \wro\  presents optimum opportunities for the study of its collisionless shocks in an astrophysical setting whose variable physical conditions of density, temperature, wind and orbital velocities and viewing angle remain under the simple control of gravity and orbital geometry.

The physical justification for classifying \wro's shocks as collisionless \citep{2005ApJ...629..482P} and therefore regulated by collective electromagnetic processes as opposed to gas collisions rests on the long mean-free-paths for Coulomb collisions that pertain in the plasmas that constitute the winds of the hot stars. Conclusive supporting evidence for a collisionless classification comes from \wro's status as a strong non-thermal synchrotron radio source \citep{1995ApJ...451..352W}. The relativistic electrons and magnetic fields required are both common properties of collisionless shock physics \citep[e.g.][]{2016RPPh...79d6901M} that do not arise in the collisional regime.

\subsection{Magnetic Conditions in the Shocks of \wro.}

 {Though}  \citet{2016RPPh...79d6901M} argue that the background upstream magnetic field plays a pivotal role in the formation and dissipation of   {collisionless} shocks, particle-in-cell simulations \citep{2008ApJ...681L..93K} show that the Weibel instability also operates in interacting unmagnetised plasmas to generate magnetic fields whose energy density amounts to 1\% or 2\% of the upstream bulk kinetic energy density. At the stagnation point in the terminal-velocity regime in that case
\begin{equation}
    \frac{1}{8\pi}B^2 = W_B \frac{1}{2} \rho V\infty^2
\end{equation}
suggesting a shock-generated field of
\begin{equation} \label{eq:WeibelB}
    B \approx 1141 \frac{1}{R_X} (W_B \dot{M}_{-6} V_8)^{\frac{1}{2}} G
\end{equation}
where $W_B$ is the fractional magnetic-field energy density and otherwise the units are those of equation~(\ref{eq:Nx}) including $R_X$ in solar radii. Shock-generated fields in initially unmagnetized media have been successfully reproduced in the laboratory \citep{2017PhPl...24d1410H} and are an alternative to the common practice following \citet{1967ApJ...148..217W} of extrapolating a stellar field into the wind, especially since strong surface fields are rare in O stars \citep{2017MNRAS.465.2432G} and undetected so far in Wolf-Rayet stars \citep{2014ApJ...781...73D}. At periastron, equation~(\ref{eq:WeibelB}) implies a field of $0.05-0.1~G$ at the stagnation point, and a shock dissipation scale length defined by the proton gyroradius of a few km. This field is about two orders of magnitude greater than those typically encountered in interplanetary space in front of the Earth's magnetosphere.

\subsection{Competitive Shock Cooling}
 
The combination of X-ray and \ion{C}{3}~$\lambda5696$ luminosities in the shock energy budget suggests that the ideal $1/D$ prediction of colliding-wind theory might still make reasonable sense for \wro. If scalable geometry also applies, the O-star shock would lie only 36~\rsun\ from the center of the star at periastron, only a few radii above the stellar surface. The evident competition between X-ray and \ion{C}{3}~$\lambda5696$ cooling in such extreme conditions could conceivably be density or radiation dependent \citep{2012A&A...545A..95M} if some other mechanism is excluded like charge exchange due to mixing between the two winds \citep{Pollock:2012qv}. However, when it is understood that lines at long wavelengths often classified as recombination lines can provide cooling under certain conditions to rival or surpass the X-ray luminosity, it will be necessary to consider other transitions such as \ion{He}{1}~$\lambda5876$ \citep{2003ApJ...596.1295M} and \ion{He}{1}~$\lambda10830$ for which \citet{2021MNRAS.503..643W} have shown the importance of extensive phase coverage. Pending these assessments, it does not yet seem necessary to appeal to the suggestion of a variable effective mass-loss rate due to clumping effects \citep{2021MNRAS.500.4837Z} to account for the unexpected decrease in \wro's X-ray luminosity near periastron. Rather, the plasma has found another way to cool.

Future observations would be able to establish more firmly the nature of the cooling competition in at least three respects: first, more complete phase coverage including templates; second, simultaneous X-ray and optical line velocity profiles; and third, how close the \ion{C}{3} emission peak is to the 2024 November 22 periastron in comparison with the 1-day flow time. These will all allow judgement of whether the rival cooling mechanisms compete on microscopic, as seems likely, or macroscopic scales.

In other WC+O binary systems of shorter period and smaller separation, WR~42, WR~48 and WR~79 \citep{2002MNRAS.335.1069H} and WR~113 \citep{2018MNRAS.474.2987H}, the competition may have been decided in favour of \ion{C}{3}~$\lambda5696$ by their luminous excesses and weak or undetected X-ray luminosities. %

\subsection{X-ray Absorption Tomography}

While the X-ray luminosity can be be considered almost an integrated thermodynamic state variable that responds to externally imposed conditions like %
an ideal astrophysical Carnot cycle, the observed absorption of the X-ray spectrum is sensitive to the morphology of whatever cold gas intervenes between source and observer. The model spectrum used to estimate the absorption as a function of phase simply incorporated a single line-of-sight column density to the stagnation point in the flow. This is adequate for the current campaign in which the scarce resource of observing time was shared between multiple observations to yield spectra of modest statistical weight at many different phases. The X-ray source is almost certainly extended on the scale of the binary separation. Though its detailed spatial form is not known, it is small enough for sharp features to appear in Figure~\ref{fig:nh} that delineate the morphology of the cool gas as the orbit rotates. Notably, the duration of the transition between O-star and WC-star winds implies that the post-shock gas occupies a significant angular size of $26\degree$ which Figure~\ref{fig:cwbsim} suggests is due in about equal measure to the angular extent of the post-shock gas and the curvature of the shock front. As a consequence, the implication is that the shock has not suffered collapse at this point close to periastron from radiative losses.

The measurements made separately near both stellar conjunctions have enabled the estimates above of the mass-loss rates of both companion stars of
$\dot{M}_{WC7}=1.7\times10^{-5}$~\msun~yr$^{-1}$ and $\dot{M}_{O5}=3.7\times10^{-7}$~\msun~yr$^{-1}$
implying a measured value of $\eta=44$ that is reasonably close to the wind-balance parameter adopted in Table~\ref{tab:params}. This looks encouraging, though there remain challenges for future, 3D dynamical models to resolve in accounting for the phase-dependent absorption through the WC-star wind where the simple absorption model of Eq.~\ref{eq:Nx} breaks down.

The simple model of the phase-dependent O-star wind absorption implies consistent mass-loss rate estimates from measurements made over more than a year. The mass-loss rate may be low and dependent on elemental abundances that take no account of any mass-exchange that may have taken place in the course of the binary system's evolution but it does have the merit of demonstrating the sensitivity of X-ray absorption measurements of colliding-wind systems for making realistic clumping-free estimates of the mass-loss rates of the O stars in these systems and determining the extent of the absorbing media.

By contrast, point estimates following the same procedure of the WC-star mass-loss rate from its absorption lead to inconsistent values depending on orbital phase that differ by a factor of a few. In the formation and dissipation mechanisms of collisionless shocks, it becomes necessary to consider instead the established importance of reflected ions as described, for example, by \cite{2015cssp.book.....B} and \citet{2021ApJ...908...40M}. Reflected ions form a ``sheath'' of cool material around the collisionless shock, and, for WR140, the sheath on the WC7-side of the shock would consist largely of He \& C rich material with large X-ray opacity.  Observations of collisionless shocks in supernovae suggest that the reflected-ion fraction increases with Mach number, reaching up to 50\% or more of the material entering the shock \citep{2013SSRv..178..633G}. The similarities between collisionless shocks in colliding-wind binaries and supernovae was emphasized  earlier \citep{2005ApJ...629..482P}.

Figure \ref{fig:WC7Nx} shows that, at the highest columns observed, there’s relatively little change of $N_x$ with $\theta/b$, indicating a thick source of absorbing material which remains relatively independent of the orbital geometry as the stars move from O star inferior conjunction through WC star inferior conjunction and beyond.  \citet{2021MNRAS.500.4837Z} used 2D hydrodynamic models of \wro's \rxte\ data precisely to argue for an additional absorption component, although leaving its origin open.  In this orbital phase range, the WC star is in front of the shock boundary, so it seems natural  to suggest that the outer surface of \wro's collisionless shocks are enveloped in a sheath of reflected ions through which the X-rays from the colliding wind shock pass and that dominates the observed absorption in the range of maximum absorption, $0.01<\theta/b< 0.02$,  shown in Figure \ref{fig:WC7Nx}. The ionization state and chemical abundance of the reflected ion  sheath would be similar to upstream WC7 wind and present an immediate additional absorption to the X-rays originating immediately on the other side of the shock discontinuity. Because X-ray absorption would occur mainly in the enriched carbon content of the WC-star material, this would be the first direct detection outside the solar system of the reflection of heavy ions in high Mach-number collisionless shocks, and an astrophysical analog to  the \ion{He}{2} reflected ion sheath in the magnetosphere directly detected by \cite{2021arXiv210610214M}. In order to satisfy mass continuity the sheath would also have some dynamical structure that involves flow parallel to the shock that might be visible in the velocity structure of long-wavelength absorption lines in either stellar continuum at favourable orbital phases as discussed below for \ion{He}{1}~$\lambda$10830. In \wro\ it also becomes possible to explore the shock formation process involving ion reflection as a function of density and velocity shear driven by the orbital motion of the stars.

\subsubsection{X-ray Absorption and Dust}

\begin{figure}[htbp] %
   \centering
   \includegraphics[width=7in]{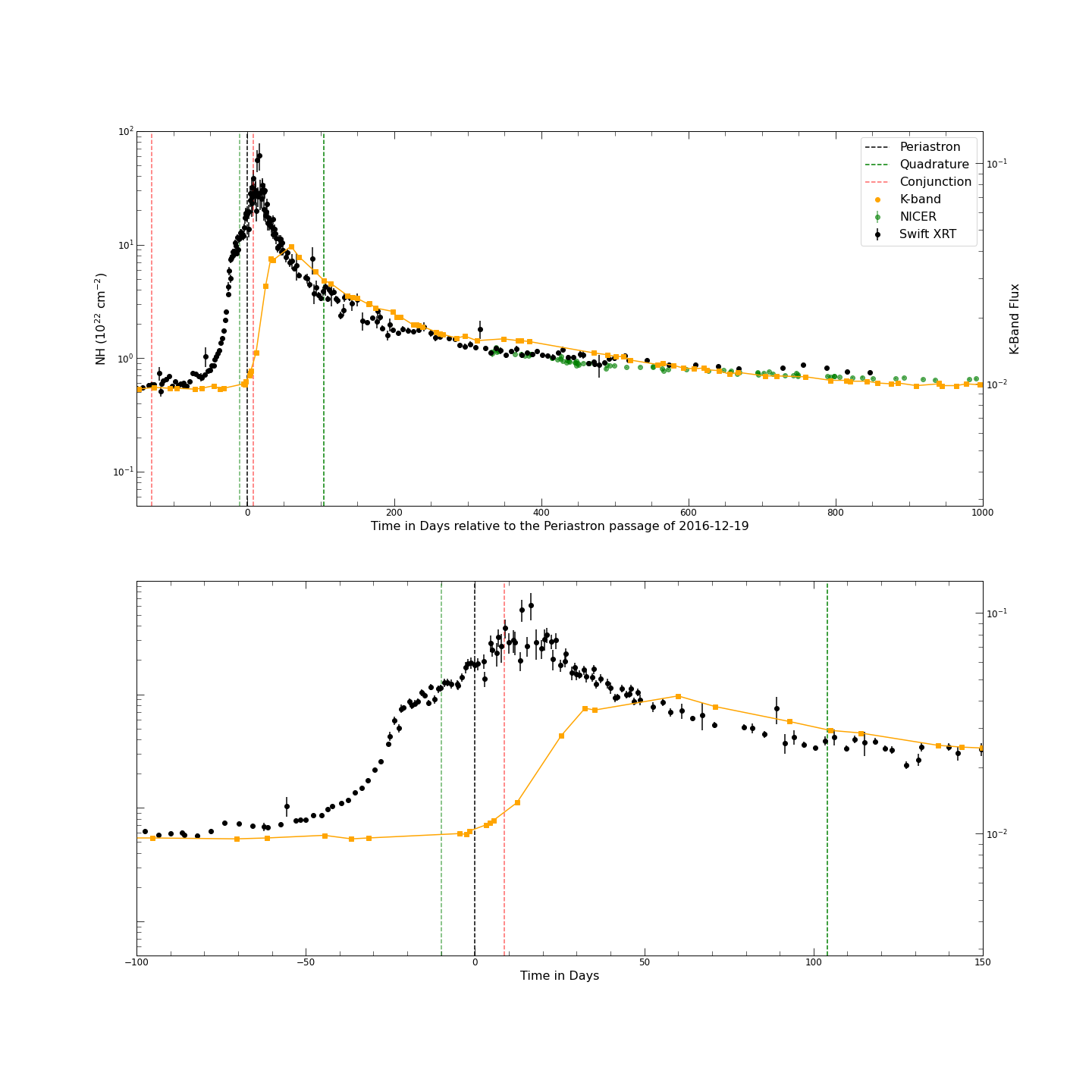} 
\caption{Comparison of the X-ray column density derived from \swift\ and \nicer\ spectra, with the variation in relative K-band flux %
vs. orbital phase around the 2016 periastron passage. The decline in $N_{H}$ is well correlated with the decline in relative K-band flux.}
   \label{fig:nh-k}
\end{figure}

We compared the column density variations to the $K$-band lightcurve in Figure~\ref{fig:nh-k} for contemporaneous data taken as part of the 2016 periastron passage observing campaign. The $K$-band lightcurve, a measure of the brightness of the periodic dust formation, is shown in orange.  The $K$-band brightness increases almost exactly at periastron passage (although it's not clear why the dust appears so precisely here and not sometime before or after), which suggests that dust nucleation begins very near  periastron or that it first becomes visible to the observer at periastron passage. In either case there's no physical reason why this coincidence would occur, unless the cold, dust-forming gas is embedded within the hot shocked gas and somehow shielded from the harsh radiative environment near the WC7 and O5 stars  \citep[as suggested by ][]{Williams:2009rf} and the X-ray emission of the shocked gas in the collision zone. %
As noted above, the rapid increase in X-ray column density begins about 50 days before periastron passage.  About one month after periastron passage, the decline in K-band brightness is (unexpectedly) very similar to the decline in column density.  The fading in $K$-band brightness is believed to be due to the cooling of dust in the dust shell as it expands away from \wro. The similarity between the decline in $K$-band brightness and decline in X-ray column may indicate a physical connection between the X-ray absorption and the dust shell, i.e. it may be that some of the X-ray absorption arises in the dust shell itself, as well as  {in} the unshocked wind of the WC star.  

In Figure  \ref{fig:nh-k}, 
the  relative $K$-band flux is derived from the $K$-band magnitudes  as $F_{K} = 10^{-K/2.5}$, where $F_{K}$ 
\rev{is}
the relative $K$-band flux and $K$ the $K$-band magnitude. The $K$-band increase starts within five days of periastron passage (unfortunately there is no $K$-band data in the interval $\approx\pm2.5$~days around periastron passage in 2016). The increase in $K$-band brightness occurs well after the start of the increase in X-ray column density, and maximum $K$-band brightness occurs after maximum X-ray absorption (which, as noted, occurs very near superior conjunction of the O star).  
The $K$-band brightness does not reach its maximum until about  70 days after periastron.  Assuming the speed of the dust is 1600~\ks \citep{1990MNRAS.243..662W}, after 70 days the dust has moved 
65 AU, about 5
times the semi-major axis of \wro. 
After $K$-band brightness reaches its maximum, it begins to decline exponentially, with an e-folding time of about 153 days. In the $H-$band and other short-wavelength IR bands, the brightness starts to decline shortly after dust formation ends %
\citep{1990MNRAS.243..662W}. At longer IR bands, the brightness maximum is reached at later orbital phases and the decline is slower, as the dust cools and the peak emission moves to longer wavelengths.

\subsubsection{X-ray and He I $\lambda$10830 Absorption}

\begin{figure}[htb!] %
   \centering
   \includegraphics[width=7in]{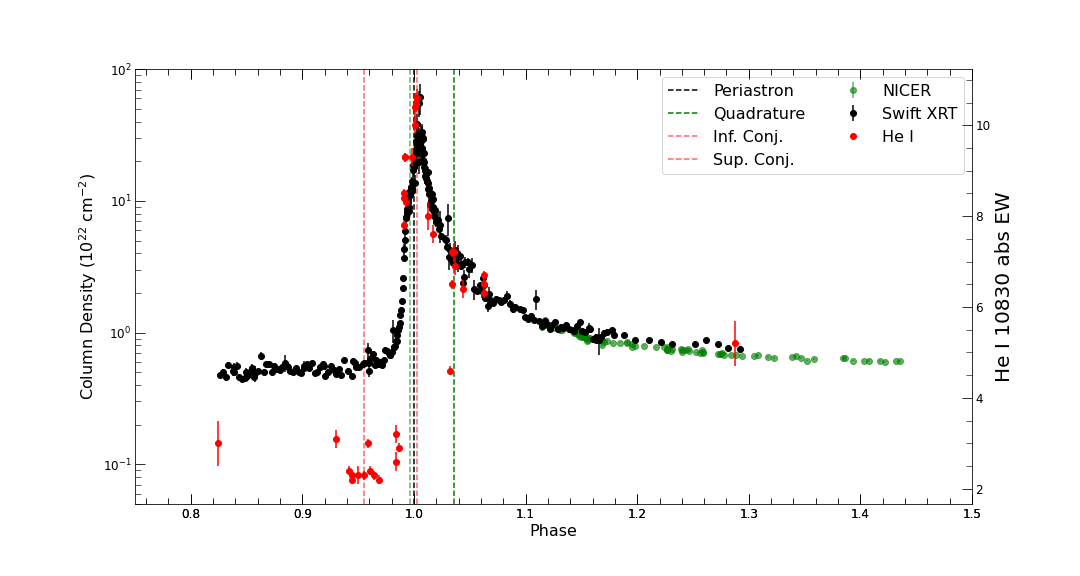} 
   \caption{Comparison of the equivalent width of the He~I 1.083$\mu$ P-Cygni absorption component from Williams et al. (2021) to the X-ray column densities as measured from the analysis of the \swift\ and \nicer\ spectra versus orbital phase.  }
   \label{fig:NH-He}
\end{figure}

\cite{2021MNRAS.503..643W} describes the variation of the P-Cygni He I  $\lambda$10830 line near the 2009 and 2016 periastron passages.  The He~I  $\lambda$10830 emission and absorption components are primarily produced by the He-rich wind of the WC7 star.  This line is both very bright, with a flux exceeding the 2~--~10 keV X-ray flux for most of the orbit and complex, showing an emission sub-peak which varies with orbital phase in strength and radial velocity.  The P-Cygni absorption is seen in the radial velocity range $-3000~\kms<V<-2000~\kms$, and so is produced by He-rich gas from the WC7 star lying along the line of sight in front of the stars. We compared the equivalent widths of the He I  $\lambda$10830 P-Cygni absorption (given in Tables 1 \& 2 of Williams et al. 2021) to the X-ray absorption column measured from the \swift\ and \nicer\ spectra (see Fig. \ref{fig:NH-He}).

As discussed by Williams et al., the observed variation in the equivalent width  of the He~I absorption is mostly due to absorption variations in the WC7 wind along the line of sight to the O star, since the amount of WC7 material in front of the O star varies systematically as the orientation and separation of the stars change with orbital motion.  There's an interval of (mostly) weak absorption near the time of inferior conjunction of the O star since near this phase O star is mostly viewed through its own lower-density wind which fills the shock cavity.  Starting at roughly $\phi\approx 0.98$ (when the trailing arm of the colliding wind bow shock passes the observer's line of sight)  both stars and the colliding wind shock  are viewed through the dense wind of the WC7 star and the X-ray column and He~I absorption increase rapidly, reach maxima near superior conjunction of the O star, and then gradually decline.  The variation in X-ray column is larger than the observed variation in He~I equivalent width; a drop by  about a factor of 50 in X-ray column corresponds to a decline of about a factor of 2 in He~I equivalent width. This may suggest that some of the absorption is filled in by high velocity, blueshifted emission.   As discussed by Williams et al., there's a spectrum showing anomalously strong He~I absorption near the inferior conjunction of the O star (near 
$\phi\approx0.96$), and anomalously weak absorption near the second quadrature (near 
$\phi\approx1.04$).  These  probably arise from stochastic density variations  {produced by} instabilities in the WC7 wind near the colliding-wind bow shock, as noted by \cite{2021MNRAS.503..643W}.

\subsection{The Relic Component}

The \swift\ data presented in Figure~\ref{fig:relic} show that the relic component can be first seen about 25 days prior to periastron passage up to 50 days after periastron passage, when absorption is high enough that the colliding wind component is a negligible contributor to the emission in the 0.4-2.0 keV band.  
The \swift\ data also show that the flux of this  component is  {not strongly} variable through this interval.
\cite{Sugawara:2015eu} suggested that this component is radiative recombination continua produced by the shocked wind as it cools (perhaps as a precursor to dust formation), but this seems unlikely given the \swift\ observations. In this 75 day interval around periastron the stellar winds have flowed out a distance 
$\sim$124
AU.
Assuming the recombining gas is entrained in the wind, it's difficult to see how  this gas could show a relatively constant X-ray brightness while moving such a large distance while experiencing large changes in density and temperature. The \swift\ observations of the relic component suggest that this is instead a soft, relatively constant source. This source has a flux of 
1.1$\times10^{-12}$~\fluxcgs, corresponding to a luminosity of 2.9$\times10^{32}$~ergs~s$^{-1}$. The X-ray flux from the embedded shocks in the O star wind should be approximately 
1.4$\times 10^{32}$~erg~s$^{-1}$, since the luminosity of the O5.5fc~star is $L_{O}\approx$
3.5$\times 10^{5} L_{\odot}$  \citep{2021arXiv210110563T}
using the canonical relation for X-ray luminosity from embedded wind shocks in O~stars, $L_{x}/L_{bol} \approx 10^{-7}$. It's unlikely that the O5.5fc star is the source of the ``relic'' component, however, since near periastron passage the star is behind the WC7 star, so that any soft emission in its wind would suffer significant, and significantly variable, absorption by the wind of the WC7 star.
Another alternative is that this emission is produced by embedded shocks within the wind of the WC7 star,  {which seems unlikely since} in general single WC7 stars are very weak X-ray sources.
The emission could be produced by a third unresolved O-type  star in the system, but this  {also} seems unlikely since the star would need to be similarly bright as the O5.5fc star.   \ec\  shows soft, constant X-ray emission from the ``Outer Debris Field'', the shell of ejecta produced by the eruption of the star in the 1840's.  Unresolved ejecta near the star from some past outburst of \wro\ could also reveal itself as the ``relic'' component, but there's no recorded outburst of \wro, so this alternative seems unlikely  {as well}.  Perhaps the most likely explanation is that this source is similar to the ``Central Constant Emission'' component seen in \ec\ \citep{2007ApJ...663..522H}, which is produced by residual hot shocked gas trapped between the distorted shock cone near periastron passage \citep{Russell:2016qy}.

\subsection{Comparison with \ec}
\label{sec:eccystp}

Long-period, eccentric massive binaries are hard to detect unless they present evidence of wind-wind collision emission. This is because the stars in these systems spend most of their time near apastron, in accordance with Kepler's second law.  
In the case of \wro, for example, it takes the component star 1220
days (about 40\% of the orbital period) to travel $\pm0.05$ in orbital phase around apastron, while around periastron the same interval in phase is covered in only 25
days (about 1\% of the orbital period).  Because stellar radial velocity variations are minimal at apastron, and photocenter variations non-existent, it's difficult or nearly impossible to identify such systems using classical radial velocity, photometric or proper-motion surveys in the absence of colliding-wind effects.  Such systems can be identified, as in the case of \wro\ and other carbon-rich ``dustars'' (like WR 125 and WR 19)  via periodic brightening in the IR caused by dust formation in the compressed wind collision near periastron.  But it usually requires a bit of patience and/or luck in order to catch such dust formation episodes as they are occurring.  The X-ray emission produced by colliding winds is a more constant tracer, since X-rays are produced through most of the orbit and since single WR stars are usually weak X-ray sources. %

Both \wro\ and \ec\  are now understood as long-period, highly-eccentric binary systems and first recognized as such based on variations  produced by their wind-wind collision regions.  In both cases, periodic variations in the IR were reported -- but the nature of these variations was rather mysterious and their ties to a putative companion star unclear. In \rev{both} \wro\ and \ec, such periodic photometric variations were  initially discussed in terms of ``shell events'' (periodic ejection of shells of material from a -- perhaps single --  star).  Early X-ray observations of the stars revealed unusually bright  X-ray flux, whose origin was also at first unclear, but which is now recognized as a clear sign of the presence of hot shocked gas produced by a wind-wind collision.  Establishing strictly periodic photometric variability though long, painstaking campaigns in the IR \citep{1987ApJ...312..807M} and radio \citep{White:1995fk}, \rev{coupled} with sporadic X-ray observations \citep{1985SSRv...40...63P, 1990MNRAS.243..662W} %
made a convincing case for the presence of another star in \wro, confirmed by subsequent observations in the radio and optical, along with high-resolution ground-based radial-velocity studies which spatially and spectrally resolved the orbital motions of the two stars (and even showed emission from the ionized gas in the bow shock).  
Similarly in the case of  \ec, observations of periodic variations in the equivalent width of the He~I $\lambda$10830  emission line over half a century, along with Paschen line variations \citep{1996ApJ...460L..49D, 1997NewA....2..107D}, and hard \citep{1982ApJ...256..530S}, variable \citep{1999ApJ...524..983I} X-ray emission provided persuasive evidence that \ec\ is also a long-period, highly eccentric binary (even though \ec's companion star has not been directly seen).

\begin{figure}[htbp] %
   \centering
   \includegraphics[width=7in]{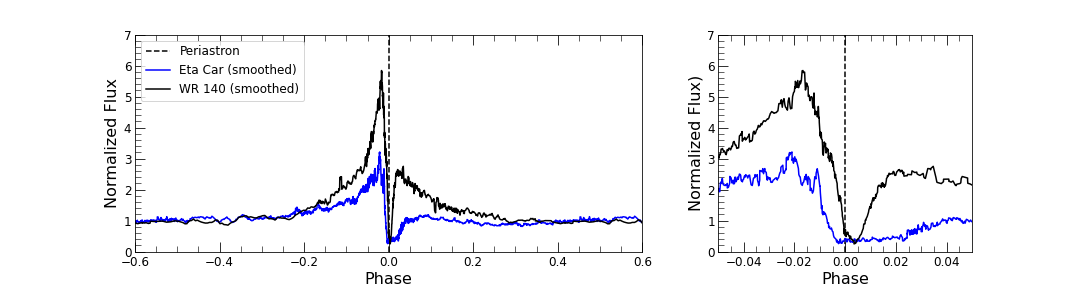} 
   \caption{\textit{Left}: Comparison of the 2--10 keV X-ray lightcurve of \wro\ from \rxte, \swift\ and \nicer, to the 2--10 keV  \rxte\ and \swift\ fluxes for \ec, normalized at apastron.  \ec\ is about an order of magnitude brighter than \wro\ at apastron.  
   The phases for \ec\ are calculated using the periastron passage epoch $T_{o}=$ JD 2,456,874.4 from \cite{2016ApJ...819..131T} and the X-ray period, 2023.7~days, and fluxes from \cite{2017ApJ...838...45C}. The lightcurves for each binary have been combined and smoothed using a 5-point boxcar for display purposes. \textit{Right}: The phase interval near periastron passages for the two binaries.
}
   \label{fig:wr-ec}
\end{figure}

The X-ray behavior of \ec\ and \wro\ shows similar variation with orbital phase -- long, slow rise in flux, a sudden drop accompanied by an increase in absorption, a brief flux minimum, the appearance of a soft component during the minimum --  but also important differences.  The 2-10 keV X-ray minimum in \ec\ appears flat-bottomed (mostly due to the presence of circumstellar X-ray background emission) \rev{and to}
last for 30--60~days before increasing in flux.  \wro's X-ray minimum state in the same band only lasts for about 10 days once the minimum is reached. 
Figure \ref{fig:wr-ec} compares the X-ray flux variations of \wro\ and \ec, normalized to their fluxes at apastron.  Although \ec\ is about an order of magnitude brighter in the 2--10 keV band (and about 15$\times$ brighter in X-ray luminosity), the  ratio of X-ray flux at maximum to X-ray flux at apastron is about a factor of two greater for \wro\ compared to \ec. %
These differences may be due to differences in orbital eccentricity, if the orbital eccentricity of \ec\ is lower than the eccentricity of \wro\ ($e=$0.90). Other important differences that shape the X-ray lightcurves are the relative terminal stellar wind speeds $V_{\infty}$ of the component stars in the binary ($V_{\infty}$ is about a factor of 8 different for the stars in \ec, and nearly the same for the two stars in \wro), along with wind (and circumstellar) absorption.

On approach to X-ray maximum, \ec's X-ray flux shows large variations on timescales of weeks to days \citep{1997Natur.390..587C, 2009ApJ...707..693M}. This variability has not been seen in any of the pre-minimum phase intervals for \wro, and flux variations (apparently stochastic) appear to be $<10-20$\% at any phase.  \cite{2009ApJ...707..693M} suggested that the flux variations in \ec\ are driven by large, stochastic overdensities  (``clumps'') in the massive, slow wind of the LBV primary, \ec-A.  If this is the case, the relative smoothness of the \wro's X-ray lightcurve may reflect the smoothness of the wind of the WC7 star, perhaps because its higher-velocity wind helps  {reduce} out any perturbations before they can grow substantially.

\input{Conclusions}

\begin{acknowledgements}

We would especially like to acknowledge the \rxte, \swift, and \nicer\ teams, for their heroic efforts in scheduling these observations.  We'd also like to acknowledge the project scientists, Jean Swank, Neil Gehrels, Brad Cenko, \rev{Keith Gendreau and Zaven Arzoumanian} for granting director's discretionary time to help define these lightcurves.  \rev{This work made use of data supplied by the UK Swift Science Data Centre at the University of Leicester.} We'd also like to thank the anonymous and unsung peer reviewers who awarded time to these observing programs. This research was supported through NASA cooperative agreement NNG06EO960A.  This research made use of the Astrophysics Data System and the HEASARC archive. We express our appreciation to  the  amateur astronomers for their dedication and herculean efforts to observe \wro\ during its 2009 periastron passage. This work has made use of data from the European Space Agency (ESA) mission {\it Gaia} (\url{https://www.cosmos.esa.int/gaia}), processed by the {\it Gaia} Data Processing and Analysis Consortium (DPAC, \url{https://www.cosmos.esa.int/web/gaia/dpac/consortium}). Funding for the DPAC has been provided by national institutions, in particular the institutions
participating in the {\it Gaia} Multilateral Agreement. A. F. J. Moffat is grateful for financial aid from NSERC (Canada). D. Espinoza gratefully acknowledges support from NASA grants  \#80NSSC19K1451 and  \#80NSSC21K0092, and SAO grant \#GO9-20015A thru NASA.
M. F. Corcoran is supported under the CRESST-II cooperative agreement  \#80GSFC17M0002 with the NASA/Goddard Space Flight Center.
C.~M.~P.~Russell  was supported by SAO grant \#GO0-21006A  through NASA; this support is gratefully acknowledged.

\end{acknowledgements}

\facility{RXTE(PCA)}, \facility{Swift(XRT)}, \facility{NICER(XTI)}.

\bibliography{mfc_wrx}

\appendix

\restartappendixnumbering

\section{Sample Spectra}

\begin{figure}[htbp] %
   \centering
   \includegraphics[width=3in]{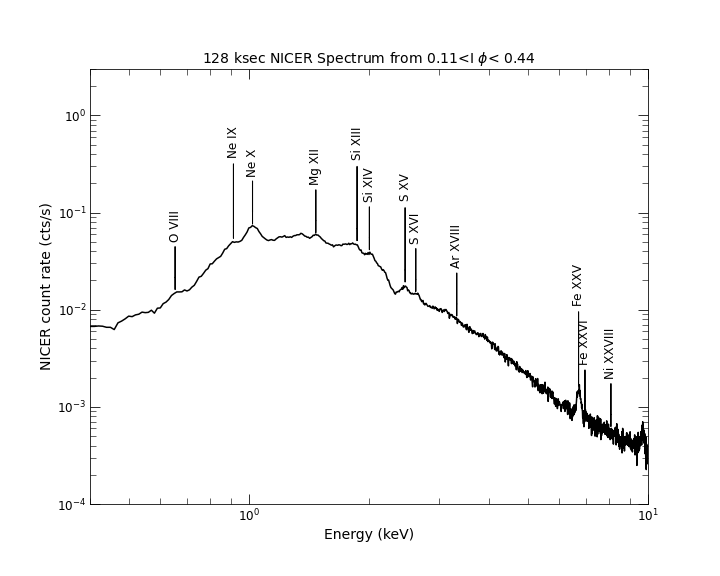} 
      \includegraphics[width=3in]{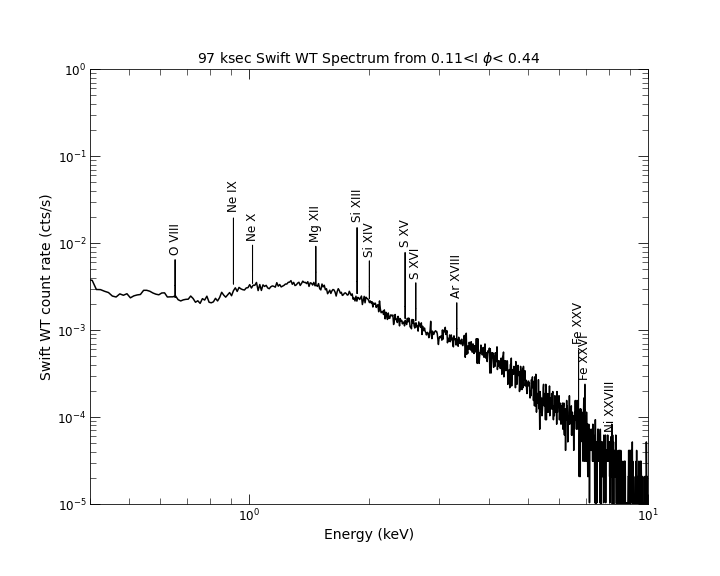} 
      \includegraphics[width=3in]{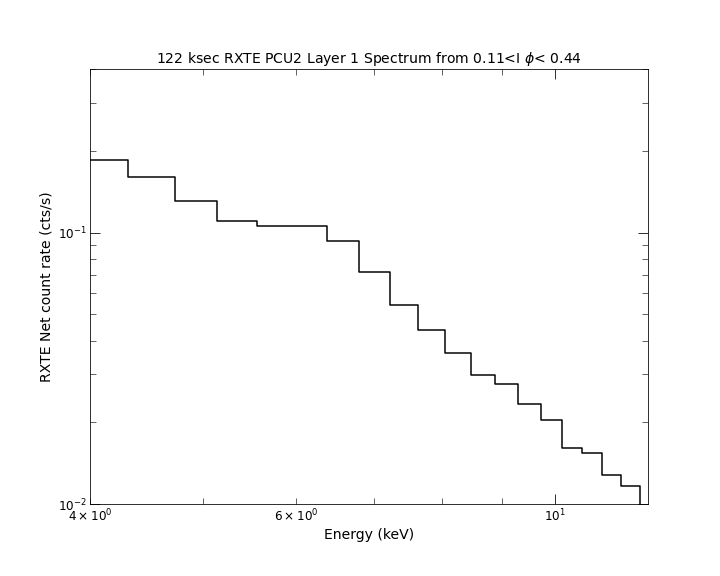} 
  \caption{
Comparison of combined spectra from the 3 instruments (\nicer-XTI, \swift-XRT and \rxte-PCA) in a common interval of WR 140's orbital phase ($0.1<\phi<0.4$). Some strong lines are marked in the \nicer-XTI and \swift-XRT spectrum (the poor spectral resolution of the \rxte-PCA neans it's incapable of resolving line features, though the Fe XXV feature does show up as an apparent ``shelf'' near 6.7 keV).  The \swift-XRT  PC-mode data are not shown because there were only 4 PC-mode observations in this phase interval.
   }
   \label{fig:combinedspec}
\end{figure}

To show the differences in spectral resolution and energy sensitivity, we show sample spectra in the orbital phase range $0.1<\phi<0.4$ in \ref{fig:combinedspec}. This phase range was chosen because it's the only range sampled by all 3 instruments, and also because there's relatively little variation in flux within this range. To improve statistics,  these spectra are combined from all the spectra for a given instrument in this phase range.  The  the \rxte-PCA spectrum is combined from net spectra, where the net spectrum has been accumulating after correcting individual \rxte-PCA spectra for estimated instrumental background.  No background correction has been applied to either the \nicer-XTI or \swift-XRT spectra.  

\startlongtable
\input{rxte_obslog_wr140_fixed}

\startlongtable
\input{swift_obslog_wr140_20200327_fixed}

\startlongtable
\input{nicer_obslog_wr140}

\end{document}

%% file: sumtab.tex
\begin{deluxetable}{cccccccc}
\tabletypesize{\scriptsize}
\tablecaption{Summary of the Observations \label{tab:sumtab}}
\tablehead{\colhead{Instr.} & \colhead{Num. Obs.} & \colhead{Tot. Expo.} & \colhead{Obs. Start} & \colhead{Obs. End} & \colhead{Min. Expo.} & \colhead{Max. Expo.} & \colhead{Min. Sep.}\\ \colhead{ } & \colhead{ } & \colhead{ksec} & \colhead{MJD} & \colhead{MJD} & \colhead{s} & \colhead{s} & \colhead{day}}
\startdata
RXTE PCA & 533 & 510 & 51887 & 55919 & 16 & 2192 & 0.55 \\
Swift PC & 145 & 313 & 54689 & 57891 & 376 & 19760 & 0.06 \\
Swift WT & 173 & 391 & 54843 & 58588 & 79 & 18648 & 0.07 \\
NICER XTI & 89 & 128 & 58074 & 59002 & 57 & 4501 & 0.05
\enddata

    \tablecomments{Tot. Expos. is the sum of the exposures for the individual observations; 
    Obs. Start, End are the start and stop of the observing campaigns in MJD; 
    Min., Max. Expo. are the minimum and maximum exposure for individual observations; 
    Min. Sep.  is the minimum separation between consecutive observations}
    
\end{deluxetable}

%% file: templ_mo.tex
\begin{deluxetable}{cccc}   
\tabletypesize{\scriptsize}
\tablecaption{X-ray Spectrum Template Model \label{tab:templ_mo} }
\tablehead{\colhead{Parameter} & \colhead{Unit} & \colhead{Initial Value} & \colhead{Type}} 
\startdata  
nH & 10$^{22}$~cm$^{-2}$ & 0.50 &  \\
kT$_1$ & keV & 3.68 & Fixed \\
EM$_1$ & 10$^{53}$~cm$^{-3}$ &  3.87 & Fixed \\
kT$_2$ & keV & 1.14 & Fixed \\
EM$_2$ & 10$^{53}$ cm$^{-3}$ & 1.42 & Fixed \\
kT$_3$ & keV & 0.41 & Fixed \\
EM$_3$ & 10$^{53}$ cm$^{-3}$ & 1.54  & Fixed \\
Velocity Shift & 1000 km s$^{-1}$ & $-1.37$ & Fixed \\
Line Broadening & 100 km s$^{-1}$ &5.78 & Fixed \\
Normalization factor &  & 1.0 &  \\ 
\hline \hline   
\multicolumn{4}{c}{\textit{Fixed Emission Abundances by number, N, and mass fraction, X}} \\ 
\hline   
   & \multicolumn{1}{r}{N/N$_\odot$} & N & X \\    
He & \multicolumn{1}{r}{ 1000} & 1.                  & 0.548 \\   
C  & \multicolumn{1}{r}{65247} & 0.242               & 0.398 \\   
N  & \multicolumn{1}{r}{    0} & 0.                  & 0.    \\    
O  & \multicolumn{1}{r}{ 1938} & $1.69\times10^{-2}$ & $3.70\times10^{-2}$ \\  
Ne & \multicolumn{1}{r}{ 3139} & $3.95\times10^{-3}$ & $1.09\times10^{-2}$ \\  
Mg & \multicolumn{1}{r}{ 1061} & $4.13\times10^{-4}$ & $1.37\times10^{-3}$ \\  
Al & \multicolumn{1}{r}{  976} & $2.95\times10^{-5}$ & $1.09\times10^{-4}$ \\  
Si & \multicolumn{1}{r}{  881} & $3.20\times10^{-4}$ & $1.23\times10^{-3}$ \\  
S  & \multicolumn{1}{r}{  713} & $1.18\times10^{-4}$ & $5.18\times10^{-4}$ \\  
Ar & \multicolumn{1}{r}{  602} & $2.24\times10^{-5}$ & $1.22\times10^{-4}$ \\  
Ca & \multicolumn{1}{r}{  374} & $8.75\times10^{-6}$ & $4.80\times10^{-5}$ \\  
Fe & \multicolumn{1}{r}{  297} & $1.42\times10^{-4}$ & $1.09\times10^{-3}$ \\  
Ni & \multicolumn{1}{r}{  385} & $7.02\times10^{-6}$ & $5.64\times10^{-5}$ \\  
\enddata                                                                      
\end{deluxetable}

%% file: Conclusions.tex
\section{Conclusions}
\label{sec:conc}

In the 66 years since the initial suggestion by Gold \citep[][p103]{1955IAUS....2...97.} that shock discontinuities occur in low-density plasmas on scales far shorter than their long Coulomb collisional mean-free-paths, collisionless shocks have become recognised as central for the understanding of gas dynamics throughout the Universe. Similarly, in the 45 years since the predictions by \citet{1976SvA....20....2P} and \citet{1976SvAL....2..138C} that Wolf-Rayet binary systems should be bright sources of X-rays from shocked gas produced by the dynamical interaction of two counterstreaming supersonic stellar-wind plasmas, observations have confirmed their high luminosities among hot-star X-ray sources. \wro\ is the preeminent example that unites these two themes for its exceptional X-ray brightness among other observational riches and the precision with which its orbital and stellar parameters are known, allowing the boundary conditions to be accurately defined under which the collisionless shocks arise.

The accumulation of nearly 1000 X-ray observations and nearly complete phase coverage have allowed some progress to be made towards a coherent understanding of the physics underlying its collisionless shocks and the various instruments that have been employed. \wro\ is bright enough in the X-ray band for statistical errors to be smaller than residual low-levels of systematic calibration errors. In order to exploit the promise of high-precision collisionless-shock physics enabled by \wro, attention has variously been necessary to instrumental features such as X-ray pile-up, optical loading, background contamination and cross-calibration. Within these limits and within the framework of a colliding-wind binary model in which the shocks are understood to be collisionless, a clear and consistent set of conclusions has emerged from campaigns with \rxte, \swift, and \nicer\ in the first 20 years of the 21st century regarding \wro's behaviour as a function of the phase of its highly eccentric 7.94-year orbit:

\begin{itemize}
    \item X-ray emission and absorption from \wro\ are repeatable with orbital phase
    \item The smooth X-ray lightcurve shows no evidence of instabilities or other short-term variability or of changes in its effective temperature
    \item Shock X-ray luminosity is inversely proportional to binary separation for most of the orbit
    \item Luminosity departures near periastron are largely restored in a switch from X-rays to excess optical line emission that probably operates on a microscopic level
    \item The shock energy budget probably also involves magnetic-field generation through the Weibel instability
    \item Foreground X-ray absorption occurs upstream in the wind terminal-velocity regimes of both WC7 and O5 stars according to phase allowing clumping-free estimates of their mass-loss rates and the interstellar absorption
    \item Features in the X-ray absorption curve arise from the geometry of the shock system
    \item Shock-reflected ions would naturally explain a substantial part of the X-ray absorption by the WC-star wind
    \item Dust-powered K-band photometry and He~I~$1.083\mu$ absorption are correlated in their own fashion with the falling X-ray absorption after periastron
    \item Residual soft X-rays revealed near periastron are less variable than the bulk shock emission and their origin remains uncertain
\end{itemize}

The full physical implications of the considerable body of accumulated data on \wro\ including the conclusions listed above will require detailed models to be built. In this, we are happy to anticipate sharing with modelers whatever data or summary parameters they may require. In the specific context of collisionless shocks, it is an open question how successful fluid models might be or whether kinetic particle-in-cell calculations are necessary or even feasible. In either case, the shock continuity equations that determine post-shock conditions on the basis of things we know probably need to face adjustments that conceivably involve, for example, reflected ions; magnetic-field generation; non-thermal particle acceleration; or cooling competition. The presence of a sheath of reflected ions could also provide an alternative dense environment cooler than shocked gas that might be worth considering as a possible site for dust formation.

As evident from \wro, the utility of complete and repeated phase coverage should encourage similar efforts in other binary systems. These might be especially worthwhile in objects for which even currently available piecemeal work has already shown to have interesting behavior such as V444~Cygni \citep{2015A&A...573A..43L} or $\gamma^2$~Velorum \citep{2004A&A...422..177S}.

Future work on \wro\ itself should immediately entail absorption measurements to close the vertical loop in Figure~\ref{fig:WC7Nx} to trace the comparatively slow transition between WC-star and O-star winds and to seek a global absorption minimum to test the reliability of the interstellar component estimated above. More important will be plans to explore further the complementary relationship established here between competitive means of plasma cooling in at least two ways: by establishing precise timing of the total radiative luminosity compared with the key orbital event of periastron; and by comparing the velocity profiles of the emission lines at short X-ray wavelengths and long optical or IR wavelengths. These will help to determine the extent to which the X-ray and excess optical emission arise in the same physical space and establish how prompt are electron and ionic heating.

The current data suggest that \wro's X-ray properties are highly predictable and that it could or should therefore serve as a calibration source both for high-energy instruments and for numerical models of shock formation. In any case, the way seems open for \wro\ to become a key reference point for the study of collisionless shocks. The space physics and astrophysics communities might join forces in devising ideas for observations and models to exploit the richness of its observational variety to advance understanding of physical processes that operate throughout the Universe.

%% file: rxte_obslog_wr140_fixed.tex
\begin{deluxetable*}{ccccccccc}
\tabletypesize{\scriptsize}
\tablecaption{Log of \textit{RXTE} Observations and Fluxes \label{tab:rxteobslog}}
\tablehead{\colhead{Sequence} & \colhead{MJD} & \colhead{Phase} & \colhead{Expo.} & \colhead{Net Rate} & \colhead{Rate Error} & \colhead{Bkg Rate} & \colhead{Flux}                      & \colhead{Error} \\
            \colhead{ }       & \colhead{ }   & \colhead{ }     & \colhead{s}     & \colhead{cts/s}    & \colhead{cts/s}      &  \colhead{cts/s}   &\colhead{10$^{-11}$ ergs/s/cm$^{2}$} & \colhead{10$^{-11}$ ergs/s/cm$^{2}$}}
\startdata
2000120902 & 51887.106 & 1.978 & 1472.0 & 6.56 & 0.11 & 12.29 & 5.19 & 0.17 \\
2000121607 & 51894.334 & 1.980 & 1440.0 & 6.65 & 0.12 & 13.02 & 5.56 & 0.18 \\
2000122305 & 51901.252 & 1.983 & 1248.0 & 6.87 & 0.12 & 12.21 & 5.67 & 0.19 \\
2000123004 & 51908.194 & 1.985 & 1440.0 & 6.71 & 0.12 & 12.91 & 5.56 & 0.18 \\
2001010608 & 51915.365 & 1.988 & 1584.0 & 7.07 & 0.11 & 12.64 & 5.15 & 0.15 \\
2001011406 & 51923.266 & 1.990 & 304.0 & 5.73 & 0.24 & 11.55 & 3.52 & 0.28 \\
2001011919 & 51928.827 & 1.992 & 1024.0 & 5.49 & 0.13 & 13.10 & 3.54 & 0.17 \\
2001012722 & 51936.930 & 1.995 & 1440.0 & 3.85 & 0.11 & 12.63 & 2.32 & 0.13 \\
\enddata
\end{deluxetable*}

%% file: swift_obslog_wr140_20200327_fixed.tex
\begin{deluxetable}{cccccccccc}
\tabletypesize{\scriptsize}
\tablecaption{Log of \textit{SWIFT} Observations and Fluxes \label{tab:rxteobslog} }
\tablehead{\colhead{Sequence} & \colhead{MJD} & \colhead{Phase} & \colhead{Expo.} & \colhead{Net Rate} & \colhead{Rate Error} & \colhead{Bkg Rate} & \colhead{Flux} & \colhead{Error} & \colhead{Mode}\\ 
\colhead{ } & \colhead{ } & \colhead{ } & \colhead{s} & \colhead{cts/s} & \colhead{cts/s} & \colhead{cts/s} & \colhead{10$^{-11}$ ergs/s/cm$^{2}$} & \colhead{10$^{-11}$ ergs/s/cm$^{2}$} & \colhead{ }}
\startdata
31251001 & 54689.231 & 2.946 & 1379.0 & 0.967 & 0.027 & 0.004 & 2.477 & 0.068 & pc \\
31251002 & 54690.330 & 2.946 & 3876.3 & 0.970 & 0.016 & 0.004 & 2.555 & 0.042 & pc \\
31251003 & 54692.603 & 2.947 & 516.5 & 0.917 & 0.042 & 0.003 & 2.395 & 0.110 & pc \\
31251004 & 54694.710 & 2.948 & 3658.2 & 0.952 & 0.016 & 0.004 & 2.470 & 0.042 & pc \\
31251005 & 54696.218 & 2.948 & 3738.4 & 0.973 & 0.016 & 0.004 & 2.594 & 0.043 & pc \\
31251006 & 54698.305 & 2.949 & 3811.1 & 0.861 & 0.015 & 0.004 & 2.484 & 0.044 & pc \\
31251007 & 54700.444 & 2.950 & 1028.0 & 0.836 & 0.029 & 0.003 & 2.678 & 0.092 & pc \\
31251008 & 54702.340 & 2.950 & 2346.8 & 1.028 & 0.021 & 0.004 & 2.817 & 0.058 & pc \\
\enddata
\end{deluxetable}

%% file: nicer_obslog_wr140.tex
\begin{deluxetable}{ccccccccc}
\tabletypesize{\scriptsize}
\tablecaption{Log of \textit{NICER} Observations and Fluxes  \label{tab:rxteobslog}}
\tablehead{\colhead{Sequence} & \colhead{MJD} & \colhead{Phase} & \colhead{Expo.} & \colhead{Net Rate} & \colhead{Rate Error} & \colhead{Bkg Rate} & \colhead{Flux} & \colhead{Error}\\ \colhead{ } & \colhead{ } & \colhead{ } & \colhead{s} & \colhead{cts/s} & \colhead{cts/s} & \colhead{cts/s} & \colhead{10$^{-11}$ ergs/s/cm$^{2}$} & \colhead{10$^{-11}$ ergs/s/cm$^{2}$}}
\startdata
1120010102 & 58073.892 & 4.115 & 146.6 & 8.70 & 0.25 & 0.44 & 1.95 & 0.08 \\
1120010103 & 58074.342 & 4.115 & 57.4 & 7.70 & 0.43 & 2.68 & 1.55 & 0.12 \\
1120010105 & 58078.770 & 4.117 & 187.9 & 7.94 & 0.22 & 1.15 & 1.78 & 0.07 \\
1120010106 & 58079.797 & 4.117 & 216.8 & 7.60 & 0.20 & 1.00 & 1.80 & 0.07 \\
1120010107 & 58080.471 & 4.117 & 2075.0 & 7.65 & 0.06 & 0.64 & 1.61 & 0.02 \\
1120010108 & 58081.565 & 4.118 & 4362.0 & 7.81 & 0.04 & 0.70 & 1.74 & 0.01 \\
1120010109 & 58082.338 & 4.118 & 4460.0 & 7.79 & 0.04 & 0.70 & 1.66 & 0.01 \\
1120010110 & 58105.784 & 4.126 & 3067.7 & 7.68 & 0.05 & 1.23 & 1.54 & 0.02 \\
\enddata
\end{deluxetable}